\documentclass[notitlepage]{article}

\usepackage{booktabs}
\usepackage[caption=false]{subfig}
\usepackage{xifthen}
\usepackage{xspace}
\usepackage{listings}
\usepackage{enumitem}
\usepackage{bm} 
\usepackage{array}
\usepackage{amssymb}
\usepackage{balance}
\usepackage{amsmath}
\usepackage{url}
\usepackage{stmaryrd}
\usepackage{fullpage}

\newtheorem{theorem}{Theorem}[section]
\newtheorem{example}[theorem]{Example}
\newtheorem{definition}[theorem]{Definition}

\date{}

\usepackage{todonotes}

\usepackage{tikz} 
\usetikzlibrary{topaths,calc,shapes,decorations.pathmorphing,arrows}

\definecolor{goodgreen}{rgb}{0.1, 0.5, 0.1}

\lstdefinelanguage{SQL}{
  basicstyle=\linespread{1.1}\small\ttfamily,
  tabsize=2,
  numbers=none,
  stringstyle=\color{white}\ttfamily,
  showspaces=false,
  showtabs=false,
  showstringspaces=false
}

\newcommand{\vecnormal}[1]{\textnormal{\bf #1}\xspace}
\newcommand{\vectext}[1]{\textnormal{\bf #1}\xspace}
\newcommand{\textvec}[1]{\textnormal{\bf #1}\xspace}
\newcommand{\nop}[1]{}

\newcommand{\db}{\mathcal{D}}
\newcommand{\Dom}{\mathsf{Dom}}
\newcommand{\Codom}{\textnormal{\bf D}\xspace}

\newcommand{\bigexists}{\mathop{\lower0.5ex\hbox{\scalebox{1.5}{\ensuremath{\exists}}}}}
\newcommand{\punto}{$\hspace*{\fill}\Box$}

\newcommand{\bigO}[1]{\mathcal{O}(#1)}

\newcommand{\sch}{\mathsf{sch}}
\newcommand{\tuple}[1]{(#1)}

\newcommand{\milos}[1]{\todo[inline,color=yellow]{\textsf{#1} \hfill \textsc{--Milos.}}}

\newcommand{\TAB}{\makebox[2.5ex][r]{}}%
\newcommand{\SPACE}{\makebox[0.75ex][r]{}}%

\newcommand{\LET}{\textbf{let}\xspace}%
\newcommand{\IN}{\textbf{in}\xspace}%
\newcommand{\IF}{\textbf{if}\xspace}%
\newcommand{\ELSE}{\textbf{else}\xspace}%
\newcommand{\RETURN}{\textbf{return}\xspace}%
\newcommand{\MATCH}{\textbf{switch}\xspace}%
\newcommand{\VIEW}[2][]{\ifthenelse{\isempty{#1}}{\mathsf{#2}\xspace}{\mathsf{#2[\mathit{#1}]}\xspace}}
\newcommand{\VPLUS}{\uplus}
\newcommand{\VPROD}{\otimes}
\newcommand{\VPRODBIG}{\bigotimes}
\newcommand{\VSUM}{\textstyle\bigoplus}
\newcommand{\VEXISTS}[1]{\bigexists_{#1}}

\newcommand{\RING}{\textnormal{\bf D}\xspace}
\newcommand{\RINGPLUS}{+}
\newcommand{\RINGPROD}{*}
\newcommand{\RINGZERO}{\bm{0}}
\newcommand{\RINGONE}{\bm{1}}

\newcommand{\X}{\bm{X}\xspace}%
\newcommand{\Th}{\bm{\theta}\xspace}%
\newcommand{\TR}[1]{#1^{\text{T}}}
\newcommand{\LRringC}{c\xspace}
\newcommand{\LRringS}{\bm{s}\xspace}
\newcommand{\LRringQ}{\bm{Q}\xspace}

\newcommand{\DF}{F-IVM\xspace}%
\newcommand{\DFONE}{F-IVM-ONE\xspace}%
\newcommand{\DBT}{DBT\xspace}%
\newcommand{\DBTRING}{DBT-RING\xspace}%
\newcommand{\IVM}{1-IVM\xspace}%
\newcommand{\SQLOPT}{SQL-OPT\xspace}%
\newcommand{\DFRE}{F-RE\xspace}%
\newcommand{\DBTRE}{DBT-RE\xspace}%

\title{Incremental View Maintenance\\ with Triple Lock Factorization Benefits\footnote{A short version of this work appeared in SIGMOD 2018.}}

\author{Milos Nikolic and Dan Olteanu\\
Department of Computer Science, University of Oxford, UK 
}

\begin{document}


\maketitle 
\begin{abstract}
We introduce F-IVM, a unified incremental view maintenance (IVM) approach for a variety of tasks, including gradient computation for learning linear regression models over joins, matrix chain multiplication, and factorized evaluation of conjunctive queries.
 
F-IVM is a higher-order IVM algorithm that reduces the maintenance of the given task to the maintenance of a hierarchy of increasingly simpler views. The views are functions mapping keys, which are tuples of input data values, to payloads, which are elements from a task-specific ring. Whereas the computation over the keys is the same for all tasks, the computation over the payloads depends on the task. F-IVM achieves efficiency by factorizing the computation of the keys, payloads, and updates.
 
We implemented F-IVM as an extension of DBToaster. We show in a range of scenarios that it can outperform classical first-order IVM, DBToaster's fully recursive higher-order IVM, and plain recomputation by orders of magnitude while using less memory.
\end{abstract}


\section{Introduction}
\label{sec:introduction}

Supporting modern applications that rely on accurate and real-time analytics  computed over large and continuously evolving databases is a challenging data management problem~\cite{LB:SIGMOD:2015}. Special cases are the classical problems of incremental view maintenance (IVM)~\cite{Chirkova:Views:2012:FTD,DBT:VLDBJ:2014} and stream query processing~\cite{abadi2005design,madden2005tinydb}. 

Recent efforts studied the problem of computing machine learning (ML) tasks over {\em static} databases. The predominant approach loo\-sely integrates the database systems with the statistical packages \cite{MADlib:2012,Rusu:2015,MLlib:JMLR:2016,Polyzotis:SIGMOD:Tutorial:17,Kumar:SIGMOD:Tutorial:17}: First, the database system computes the input to the statistical package by joining the database relations. It then exports the join result to the statistical package for training ML models. This approach precludes real-time analytics due to the expensive export/import steps. 
Morpheus and F push the ML task inside the database and learn ML models over static normalized data. Morpheus decomposes the task of learning generalized linear models into subtasks that are pushed down past a key-foreign key join~\cite{KuNaPa15}. F decomposes the task of learning classification and regression models over arbitrary joins into factorized computation of aggregates over joins and fixpoint computation of model parameters~\cite{SOC:SIGMOD:2016,ANNOS:PODS:2018}. This factorization may significantly lower the complexity by avoiding the computation of Cartesian products lurking within joins~\cite{BKOZ:PVLDB:2013,Olteanu:FactBounds:2015:TODS}. Real-time analytics would greatly benefit from such a tight integration of the two systems.

In this paper, we introduce \DF, a unified IVM approach for analytics over normalized data. Analytical tasks are expressed as views on joins with group-by aggregates over relations that map keys to payloads. We exemplify the power of \DF for matrix chain multiplication, factorized evaluation of conjunctive queries, and gradient computation used for learning linear regression models. Although these applications achieve different outcomes, they only differ in the specification of the sum and product operations over payloads. These payload operations can be: arithmetic addition and multiplication for queries with joins and group-by aggregates; relational union and join for listing and factorized representation of conjunctive query results; matrix addition and multiplication for gradient computation; in general, sum and product in an appropriate ring. The mechanisms for maintenance and computation over keys stay the same. \DF is thus highly extensible: efficient maintenance for new analytics over normalized data is readily available as long as they come with appropriate sum and product operations.

\DF has two ingredients. First, it leverages higher-order IVM to reduce the maintenance of the input view to the maintenance of a tree of simpler views. In contrast to classical (first-order) IVM, which does not use extra views and computes changes in the query result on the fly, \DF can tremendously speed up the maintenance task and lower its complexity by using carefully chosen extra views. Nevertheless, \DF may use substantially fewer and cheaper views than  fully-recursive IVM, which is the approach taken by the state-of-the-art IVM system DBToaster~\cite{DBT:VLDBJ:2014}. 

Second, it is the first approach to employ factorized computation for three aspects of incremental maintenance for queries with aggregates and joins: (1) it exploits insights from query evaluation algorithms with best known complexity and optimizations that push aggregates past joins~\cite{BKOZ:PVLDB:2013,Olteanu:FactBounds:2015:TODS,FAQ:PODS:2016}; (2) it can process bulk updates expressed as low-rank decompositions~\cite{TensorDecomp:2009,TensorDecomposition:2017}; and (3) it can maintain compressed representation of query results. 

\DF is implemented on top of DBToaster's open-source backend, which can generate optimized C++ code from high-level update trigger specifications. In a range of applications, \DF outperforms DBToaster and classical first-order IVM by up to two orders of magnitude in both time and space.

\begin{example}\em\label{ex:sql_sum_aggregate_intro}
Consider the following SQL query over a database $\db$ with relations $R(A,B)$, $S(A,C,E)$, and $T(C,D)$:
\begin{lstlisting}[language=SQL, mathescape, columns=fullflexible]
        Q := SELECT S.A, S.C, SUM(R.B * T.D * S.E) 
             $\,$FROM   R NATURAL JOIN S NATURAL JOIN T
             $\,$GROUP BY S.A, S.C;
\end{lstlisting}
Let us consider two different evaluation strategies for this query. 
A na\"{i}ve approach first computes the join result and then the aggregate. This can take time cubic in the size of $\db$.
An alternative strategy exploits the distributivity of the {\tt SUM} operator over multiplication to partially push the aggregate past joins and later combine these partial aggregates to produce the query result. For instance, one such partial sum over $S$ can be expressed as the view V$_\texttt{S}$:
\begin{lstlisting}[language=SQL, mathescape, columns=fullflexible] 
        V$_\texttt{S}$ := SELECT A,$\;$C,$\;$SUM(E)$\;$AS$\;$S$_\texttt{E}$ FROM S GROUP$\;$BY A,$\;$C;
\end{lstlisting}
In the view V$_\texttt{S}$, we distinguish keys, which are tuples over $(A,C)$, and payloads, which are aggregate values S$_\texttt{E}$. 
Similarly, we can compute partial sums over \texttt{R} and \texttt{T} as views V$_\texttt{R}$ and V$_\texttt{T}$. These views are joined as depicted by the {\em view tree} in Figure~\ref{fig:view_tree_sql}, which is akin to a query plan with aggregates pushed past joins. This view tree computes the query result is time linear in the size of $\db$.

\begin{figure}[t]
  \centering   
  \includegraphics[width=0.55\columnwidth]{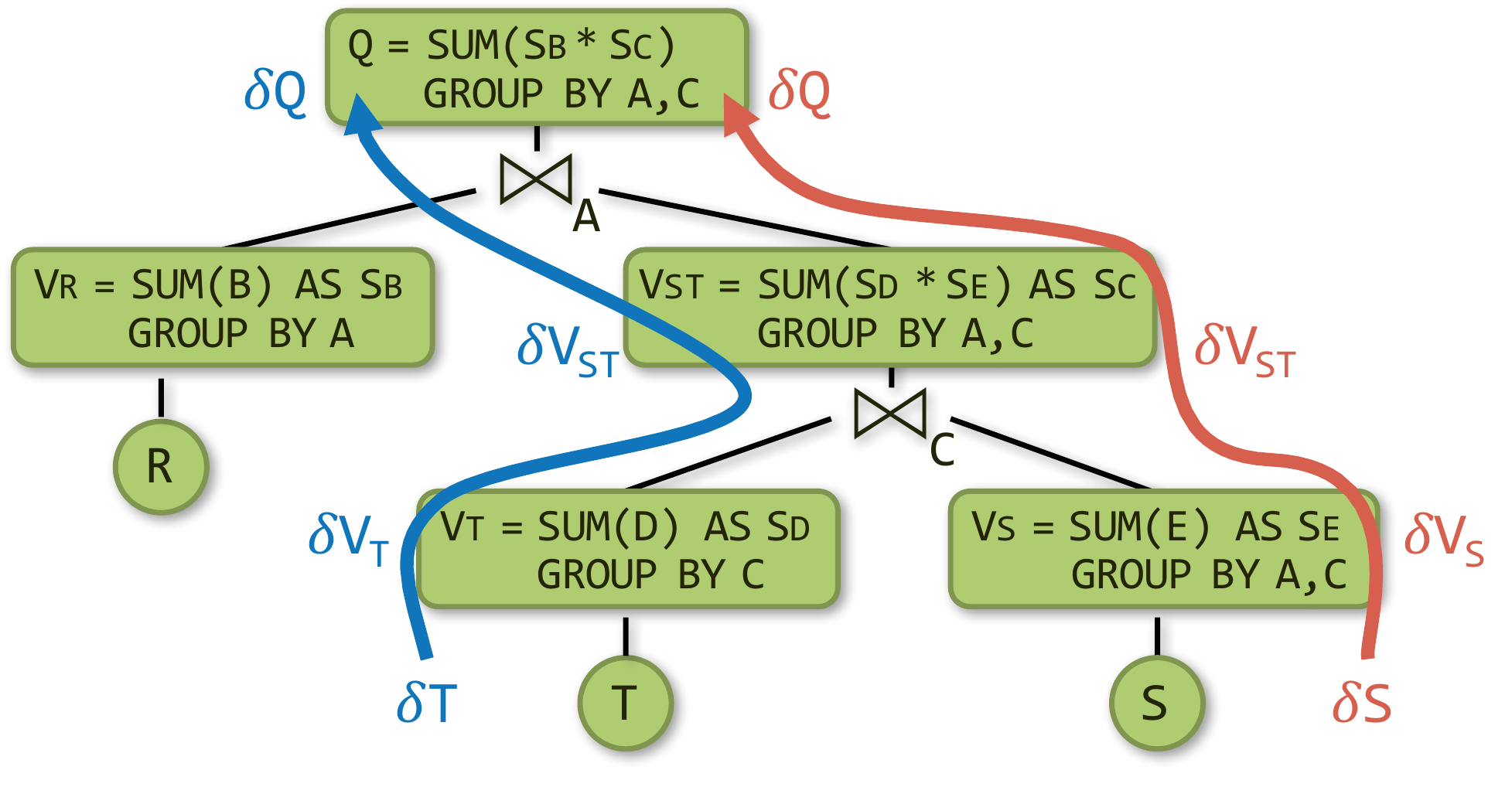}
  \caption{View tree for the query in Example~\ref{ex:sql_sum_aggregate_intro}. The delta propagation paths for updates to $S$ (right red) and to $T$ (left blue).}
  \label{fig:view_tree_sql}
\end{figure}

Consider the problem of learning a linear function $f$ with parameters $\theta_{0}$, $\theta_{D}$ and $\theta_{E}$ that predicts the label $B$ given the features $D$ and $E$, where the training dataset is the natural join of our relations:
\begin{align*}
f(D, E) = \theta_{0} + \theta_{D} \cdot D + \theta_{E} \cdot E
\end{align*}
Our insight is that the same above view tree can also compute the gradient vector used for learning $f$! The only needed adjustment is the replacement of the SQL \texttt{SUM} and \texttt{*} operators with appropriate new sum and product operators. We next explain the case of one model $f$ for each pair of values $(A,C)$; the case of one model for the entire dataset is similar. 

As detailed in Section~\ref{sec:application-lr}, the gradient of the square loss objective function requires the computation of three types of aggregates: the scalar $\LRringC$ that is the count aggregate \texttt{SUM(1)}; the vector $\LRringS$ of linear aggregates \texttt{SUM(B)}, \texttt{SUM(D)}, and \texttt{SUM(E)}; and the matrix $\LRringQ$ of quadratic aggregates \texttt{SUM(B*D)}, \texttt{SUM(B*E)}, \texttt{SUM(D*D)}, \texttt{SUM(D*E)}, and \texttt{SUM(E*E)}. These aggregates represent sufficient statistics to capture the correlation between the features and the label.
If we compute them  over each $(A,C)$ group, then we learn one model for each such group; if we compute them over the entire dataset, we then learn one model only. 

We treat these aggregates as one compound aggregate $(\LRringC,\LRringS,\LRringQ)$ so we can share computation across them. This compound aggregate can be partially pushed past joins similarly to the SUM aggregate discussed before. Its values are carried in the payloads of keys of the views in the view tree from Figure~\ref{fig:view_tree_sql}. For instance, the partial compound aggregate $(\LRringC_\texttt{T},\LRringS_\texttt{T},\LRringQ_\texttt{T})$ at the view V$_\texttt{T}$ computes for each $C$-value the count, sum, and sum of squares of the $D$-values in $T$. Similarly, the partial aggregate $(\LRringC_\texttt{S},\LRringS_\texttt{S},\LRringQ_\texttt{S})$ at the view V$_\texttt{S}$ computes for each pair $(A,C)$ the count, sum, and sum of squares of $E$-values in $S$. In the view V$_\texttt{ST}$ that is the join of V$_\texttt{T}$ and V$_\texttt{S}$, each key $(a,c)$ is associated with the multiplication of the payloads for the keys $c$ in V$_\texttt{T}$ and $(a,c)$ in V$_\texttt{S}$. This multiplication is however different from SQL's \texttt{*} as it works on compound aggregates: The scalar $\LRringC_\texttt{ST}$ is the arithmetic multiplication of $\LRringC_\texttt{T}$ and $\LRringC_\texttt{S}$; the vector of linear aggregates $\LRringS_\texttt{ST}$ is the sum of the scalar-vector products $\LRringC_\texttt{T}\LRringS_\texttt{S}$ and $\LRringC_\texttt{S}\LRringS_\texttt{T}$; finally, the matrix 
$\LRringQ_\texttt{ST}$ of quadratic aggregates is the sum of the scalar-matrix products $\LRringC_\texttt{T}\LRringQ_\texttt{S}$ and $\LRringC_\texttt{S}\LRringQ_\texttt{T}$, and of the outer products of the vectors $\LRringS_\texttt{T}$ and the transpose of $\LRringS_\texttt{S}$ and also of $\LRringS_\texttt{S}$ and the transpose of $\LRringS_\texttt{T}$. Our approach significantly shares the computation across the aggregates: The scalar aggregates are used to scale up the linear and quadratic ones, while the linear aggregates are used to compute the quadratic ones.

We now turn to incremental view maintenance. \DF operates over view trees. Whereas for non-incremental computation we only materialize the top view in the tree and the input relations, for incremental computation we may materialize additional views to speed up the maintenance task. Our approach is an instance of higher-order delta-based IVM, since an update to one relation may trigger the maintenance for several views. 

Figure~\ref{fig:view_tree_sql} shows the leaf-to-root paths taken to maintain the view result under changes to $\texttt{S}$ and to $\texttt{T}$. For updates $\delta{\texttt{S}}$ to $\texttt{S}$, each delta view $\delta{V_\texttt{S}}$, $\delta{V_\texttt{ST}}$, and $\delta{\texttt{Q}}$, is computed using delta rules:
%
\begin{lstlisting}[language=SQL, mathescape, columns=flexible] 
         $\delta$V$_\texttt{S}$ := SELECT A,$\;$C,$\;$SUM(E)$\;$AS$\;$S$_\texttt{E}$ FROM $\delta$S GROUP$\;$BY A,$\;$C;
       $\;\;\delta$V$_\texttt{ST}$ := SELECT A,$\;$C,$\;$SUM(S$_\texttt{D}$$\,$*$\,$S$_\texttt{E}$)$\;$AS$\;$S$_\texttt{C}$ 
                 FROM V$_\texttt{T}$ NATURAL JOIN $\delta$V$_\texttt{S}$ GROUP$\;$BY A,$\;$C;
          $\delta$Q := SELECT A,$\;$C,$\;$SUM(S$_\texttt{B}$$\,$*$\,$S$_\texttt{C}$)
                 FROM V$_\texttt{R}$ NATURAL JOIN $\delta$V$_\texttt{ST}$ GROUP$\;$BY A,$\;$C;
\end{lstlisting}
%
The update $\delta{\texttt{S}}$ may consist of both inserts and deletes, which are encoded as keys with positive and respectively negative payloads. In our example, a negative payload is -1 for the SUM aggregate and $(-1, {\bf 0}_{5 \times 1}, {\bf 0}_{5 \times 5})$ for the compound aggregate, where ${\bf 0}_{n \times m}$ is the $n$-by-$m$ matrix with all zero values.

\DF materializes and maintains views depending on the update workload. For updates to all input relations, it materializes the view at each node in the view tree. For updates to \texttt{R} only, it only materializes V$_\texttt{ST}$; for updates to \texttt{S} only, it materializes V$_\texttt{R}$ and V$_\texttt{T}$; for updates to \texttt{T} only, it materializes  V$_\texttt{R}$ and V$_\texttt{S}$. \DF takes constant time for updates to \texttt{S} and linear time for updates to \texttt{R} and \texttt{T}; these complexities are in the number of distinct keys in the views.
In contrast, the first-order version of \DF does not create extra views and takes linear time for updates to any of the three relations. The fully-recursive version of \DF would further materialize the join of V$_\texttt{R}$ and V$_\texttt{S}$ as it may consider more than one view tree. 

The above analysis holds for our query with one SUM aggregate. For the learning example with the nine SUM aggregates, \DF still needs the same views. Classical first-order IVM algorithms would however need to compute a distinct delta query for each of the nine aggregates -- nine in total for updates to any of the three relations. DBToaster, which is the state-of-the-art fully recursive IVM, would need to compute 28 views, nine top views and 19 auxiliary ones. Whereas \DF shares the computation across the nine aggregates, the classical IVM and DBToaster do not. This significantly widens the performance gap between \DF and its competitors.
\end{example}

\section{Data Model and Query Language}
\label{sec:preliminaries}

We recall the definition of rings. 
\begin{definition}\em
A ring $(\RING, \RINGPLUS, \RINGPROD, \RINGZERO, \RINGONE)$ is a set $\RING$ with closed binary operations $\RINGPLUS$ and $\RINGPROD$, the additive identity $\RINGZERO$, and the multiplicative identity $\RINGONE$ satisfying the axioms ($\forall a,b,c\in\RING$):
\begin{enumerate}
    \item $a \RINGPLUS b = b\RINGPLUS a$.
    \item $(a \RINGPLUS b)\RINGPLUS c = a \RINGPLUS (b \RINGPLUS c)$.
    \item $\RINGZERO \RINGPLUS a = a \RINGPLUS \RINGZERO = a$.
    \item $\exists -a \in \RING: a \RINGPLUS (-a) = (-a) \RINGPLUS a = \RINGZERO$.
    \item $(a \RINGPROD b) \RINGPROD c = a \RINGPROD (b \RINGPROD c)$.
    \item $a \RINGPROD \RINGONE = \RINGONE * a = a$.
    \item $a \RINGPROD (b \RINGPLUS c) = a \RINGPROD b \RINGPLUS a \RINGPROD c$ and $(a \RINGPLUS b) \RINGPROD c = a \RINGPROD c \RINGPLUS b \RINGPROD c$.
\end{enumerate}
A semiring $(\RING, \RINGPLUS, \RINGPROD, \RINGZERO, \RINGONE)$ satisfies all of the above properties except the additive inverse property and adds the axiom $\RINGZERO \RINGPROD a = a \RINGPROD \RINGZERO = \RINGZERO$.
A (semi)ring for which $a \RINGPROD b = b \RINGPROD a$ is commutative. 
\end{definition}

\begin{example}\em
The number sets $\mathbb{Z}$, $\mathbb{Q}$, $\mathbb{R}$, and $\mathbb{C}$ with arithmetic operations $+$ and $\cdot$ and numbers $0$ and $1$ form commutative rings. The set $\mathcal{M}$ of $(n \times n)$ matrices  forms a non-commutative ring $(\mathcal{M}, \cdot, +, 0_{n,n}, I_{n})$, where $0_{n,n}$ and $I_{n}$ are the zero matrix and the identity matrix of size $(n \times n)$. The set $\mathbb{N}$ of natural numbers  is a commutative semiring but not a ring because it has no additive inverse. Further examples are the max-product semiring $(\mathbb{R}_{+}, \max, \times, 0, 1)$, the Boolean semiring $(\{ \text{true}, \text{false} \}, \lor, \land, \text{false}, \text{true})$,  and the set semiring $(2^{U}, \cup, \cap, \emptyset, U)$ of all possible subsets of a given set $U$.\punto
\end{example}

\medskip

{\bf Data Model.}
A schema $\mathcal{S}$ is a set of variables (or attributes). For a variable $X\in\mathcal{S}$, let $\Dom(X)$ denote its domain. A tuple $\vecnormal{t}$ over schema $\mathcal{S}$ has the domain $\Dom(\mathcal{S}) = \prod_{X \in \mathcal{S}}{\Dom(X)}$. The empty tuple $\tuple{}$ is the tuple over the empty schema.

\nop{
A ring $(\RING, \RINGPLUS, \RINGPROD, \RINGZERO, \RINGONE)$ is a set with two binary operations, $+$ and $*$, which generalize the arithmetic operations of addition and multiplication, the additive identity $\RINGZERO$, the multiplicative identity $\RINGONE$, and an additive inverse for each element of $\RING$ .
Examples of rings are $\mathbb{Z}$, $\mathbb{Q}$, $\mathbb{R}$, $\mathbb{R}^2$, $\mathbb{R}^3$, and matrix ring.
}

Let $(\RING, \RINGPLUS, \RINGPROD, \RINGZERO, \RINGONE$) be a ring. A {\em relation} $\VIEW{R}$ over schema $\mathcal{S}$ and the ring $\RING$ is a function $\VIEW{R}: \Dom(\mathcal{S}) \to \Codom$ mapping tuples over schema $\mathcal{S}$ to values in $\Codom$ such that $\VIEW{R}[\vecnormal{t}] \neq \RINGZERO$ for finitely many tuples $\vecnormal{t}$. The tuple $\vecnormal{t}$ is called a {\em key}, while its mapping $\VIEW{R}[\vecnormal{t}]$ is the {\em payload} of $\vecnormal{t}$ in $\VIEW{R}$. We use $\sch(\VIEW{R})$ to denote the schema of $\VIEW{R}$. 
The statement $\vecnormal{t} \in \VIEW{R}$ tests if $\VIEW{R}[\vecnormal{t}] \neq \RINGZERO$. The size $|\VIEW{R}|$ of $\VIEW{R}$ is the size of the set $\{ \vecnormal{t} \mid \vecnormal{t} \in \VIEW{R} \}$, which consists of all keys with non-$\RINGZERO$ payloads. 
We materialize $\VIEW{R}$ as a hash map or a multidimensional array.
A database $\db$ is a collection of relations over the same ring. Its size $|\db|$ is the sum of the sizes of its relations. 
This data model is in line with prior work on $K$-relations over provenance semirings~\cite{Green:2007:ProvenanceSemirings}, generalized multiset relations~\cite{Koch:Ring:2010:PODS}, and factors over semirings~\cite{FAQ:PODS:2016}.


\medskip

{\bf Query Language.}
We consider incremental maintenance for queries with joins and group-by aggregates:
\begin{lstlisting}[language=SQL,mathescape]
        SELECT $X_1$, ..., $X_f$, SUM$(g_{f+1}(X_{f+1}) * ... * g_{m}(X_{m}))$
        FROM $R_1$ NATURAL JOIN ... NATURAL JOIN $R_n$
        GROUP BY $X_1$, ..., $X_f$
\end{lstlisting}
The group-by variables $X_1,\ldots,X_f$ are also called {\em free}, while the other variables $X_{f+1},\ldots,X_m$ are {\em bound}. The values of the {\tt SUM} aggregate are from a ring $(\RING, \RINGPLUS, \RINGPROD, \RINGZERO, \RINGONE)$. The lifting functions $g_{k}: \Dom(X_k) \to \RING$, for $f < k \leq m$, map (lift) variable values to elements in $\RING$. The \texttt{SUM} operator uses the addition $\RINGPLUS$ from $\RING$. More complex aggregates can be expressed using the sum and product operations from the ring. 

Instead of the SQL notation, we use the following encoding:
\begin{align*}
  \VIEW[X_1,\ldots,X_f]{Q} = \VSUM_{X_{f+1}} \cdots \VSUM_{X_{m}} \VPRODBIG_{i \in [n]} \VIEW[\mathcal{S}_i]{R_i},
\end{align*}
where $\VPROD$ is the join operator, $\VSUM_{X_{f+1}}$ is the aggregation operator that  marginalizes over the variable $X_{f+1}$, and each relation $\VIEW{R_i}$ maps keys over schema $\mathcal{S}_i$ to payloads in $\RING$. We also need a union operator $\VPLUS$ to express updates (insert/delete) to relations.

Given a ring $(\RING, \RINGPLUS, \RINGPROD, \RINGZERO, \RINGONE)$, relations $\VIEW{R}$ and $\VIEW{S}$ over schema $\mathcal{S}_1$ and relation $\VIEW{T}$ over schema $\mathcal{S}_2$, a variable $X \in \mathcal{S}_1$, and a lifting function $g_{X}:\Dom(X) \to \RING$, we define the operators as follows:
\\[1pt]
\scalebox{1}{
\begin{tabular}{@{}l@{~~~~~}r@{~}r@{\;}l@{}}
\multicolumn{4}{@{~}l}{\em union:}\\
& \hspace*{2em} $\forall\vecnormal{t} \in \mathsf{D}_1{:}$ & 
  $(\VIEW{R} \VPLUS \VIEW{S})[\vecnormal{t}]$ & 
  $= \VIEW{R}[\vecnormal{t}] + \VIEW{S}[\vecnormal{t}]$ \\[4pt]

\multicolumn{4}{@{~}l}{\em join:}\\
& $\forall\vecnormal{t} \in \mathsf{D}_2{:}$ & 
  $(\VIEW{S} \VPROD \VIEW{T})[\vecnormal{t}]$ & 
  $= \VIEW{S}[\pi_{\mathcal{S}_1}(\vecnormal{t})] * \VIEW{T}[\pi_{\mathcal{S}_2}(\vecnormal{t})]$\\[4pt]

\multicolumn{4}{@{~}l}{\em aggregation by marginalization:}\\[4pt]
& $\forall\vecnormal{t} \in \mathsf{D}_3{:}$ & 
  $(\VSUM_{X} \VIEW{R})[\vecnormal{t}]$ & 
  $= \textstyle\sum \,\{\, \VIEW{R}[\vecnormal{t}_1] \,\RINGPROD\, g_{X}(\pi_{\{X\}}(\vecnormal{t}_1)) \mid \vecnormal{t}_1 \in \mathsf{D}_1, \vecnormal{t} = \pi_{\mathcal{S}_1 \setminus \{X\}}(\vecnormal{t}_1) \}$\\
\end{tabular}
}\\[6pt]
where $\mathsf{D}_1 = \Dom(\mathcal{S}_1)$, $\mathsf{D}_2 = \Dom(\mathcal{S}_1 \cup \mathcal{S}_2)$, $\mathsf{D}_3 = \Dom(\mathcal{S}_1 \setminus \{X\})$, and $\pi_{\mathcal{S}}(\vecnormal{t})$ is a tuple representing the projection of tuple $\vecnormal{t}$ on the schema $\mathcal{S}$.

\begin{example}\em
Consider the relations over a ring $(\RING, \RINGPLUS, \RINGPROD, \RINGZERO, \RINGONE)$:
\begin{center}
  \small
  \begin{tabular}{lll}
    \begin{tabular}[t]{@{}l@{~~}l@{~$\to$~}l}
      $A$ & $B$ & $\VIEW{R}[A,B]$ \\\toprule
      $a_1$ & $b_1$ & $r_1$ \\
      $a_2$ & $b_1$ & $r_2$ \\
    \end{tabular}
    &
    \begin{tabular}[t]{@{}l@{~~}l@{~}l@{~}l}
      A & B & $\to$ & $\VIEW[A,B]{S}$ \\\toprule
      $a_2$ & $b_1$ & $\to$ & $s_1$ \\
      $a_3$ & $b_2$ & $\to$ & $s_2$ 
    \end{tabular}
    &
    \begin{tabular}[t]{@{}l@{~~}l@{~}l@{~}l}
      B & C & $\to$ & $\VIEW[B,C]{T}$\\\toprule
      $b_1$ & $c_1$ & $\to$ & $t_1$ \\
      $b_2$ & $c_2$ & $\to$ & $t_2$ 
    \end{tabular}
  \end{tabular}
  \end{center}
  The values $r_1$, $r_2$, $s_1$, $s_2$, $t_1$, $t_2$ are non-$\RINGZERO$ values from $\RING$. 
  The operators $\VPLUS$, $\VPROD$, and $\oplus$ are akin to classical union, join, and aggregation:\\[-9pt]
\begin{center}
  \small
  \begin{tabular}{ll}
    \begin{tabular}[t]{@{}l@{~~}l@{~}l@{~}l}
      A & B & $\to$ & $(\VIEW{R} \VPLUS \VIEW{S})[A,B]$ \\\toprule
      $a_1$ & $b_1$ & $\to$ & $r_1$ \\
      $a_2$ & $b_1$ & $\to$ & $r_2+s_1$ \\
      $a_3$ & $b_2$ & $\to$ & $s_2$
    \end{tabular}
    &
    \begin{tabular}[t]{@{}l@{~~}l@{~~}l@{~}l@{~}l}
      A & B & C & $\to$ & $\big((\VIEW{R} \VPLUS \VIEW{S}) \VPROD \VIEW{T}\big)[A,B,C]$ \\\toprule
      $a_1$ & $b_1$ & $c_1$ & $\to$ & $r_1*t_1$ \\
      $a_2$ & $b_1$ & $c_1$ & $\to$ & $(r_2 + s_1)*t_1$ \\
      $a_3$ & $b_2$ & $c_2$ & $\to$ & $s_2*t_2$
    \end{tabular}
  \end{tabular}\\[4ex]
  \begin{tabular}[t]{@{}l@{~~}l@{~}l@{~}l}
      B & C & $\to$ & $\big(\VSUM_{A}(\VIEW{R} \VPLUS \VIEW{S}) \VPROD \VIEW{T} \big)[B,C]$ \\\toprule
      $b_1$ & $c_1$ & $\to$ & $r_1 * t_1  \,*\, g_{A}(a_1) + (r_2 + s_1) * t_1  \,*\, g_{A}(a_2)$ \\
      $b_2$ & $c_2$ & $\to$ & $s_2 * t_2 \,*\, g_{A}(a_3)$
    \end{tabular}
  \end{center}
 where $g_A: \Dom(A) \to \RING$ is the given lifting function for $A$.
  \punto
\end{example}

\begin{example}\em
\label{ex:sql_count}
We show how to encode the following SQL query over tables $R(A,B)$, $S(A,C,E)$, and $T(C,D)$ into our formalism:

\begin{lstlisting}[language=SQL,columns=flexible]
      Q = SELECT SUM(1) FROM R NATURAL JOIN S NATURAL JOIN T;
\end{lstlisting}

Assuming the ring $\mathbb{Z}$, we encode the table $R$ as a relation $\VIEW{R}: \Dom(A) \times \Dom(B) \to \mathbb{Z}$ that maps tuples $(a,b)$ to their multiplicity in $R$; similarly, we encode the tables $S$ and $T$ as relations $\VIEW{S}$ and $\VIEW{T}$. Our SQL query is then translated into: 
\begin{align*}
  \VIEW[~]{Q} = \VSUM_{A}\VSUM_{B}\VSUM_{C}\VSUM_{D}\VSUM_{E} \VIEW[A,B]{R} \VPROD \VIEW[A,C,E]{S} \VPROD \VIEW[C,D]{T}
\end{align*}
where the lifting functions used in marginalization map all values to $1$.
Remember that by definition $\VIEW{R}$, $\VIEW{S}$, and $\VIEW{T}$ are finite relations.
The relation $\VIEW{Q}$ maps the empty tuple $()$ to the count. 
\punto
\end{example}

\begin{example}\em\label{ex:sql_sum_aggregate}
Let us consider the SQL query from Example~\ref{ex:sql_sum_aggregate_intro}, which computes \texttt{SUM(R.B$\,$*$\,$T.D$\,$*$\,$S.E)}, and assume that $B$, $D$, and $E$ take values from $\mathbb{Z}$.
We model the tables $R$, $S$, and $T$ as relations mapping tuples to their multiplicity, as in Example~\ref{ex:sql_count}. 
The variables $A$ and $C$ are free while  $B$, $D$, and $E$ are bound. When marginalizing over the bound variables, we apply the same lifting function to all variables: $\forall x \in \mathbb{Z}: g_{B}(x) = g_{D}(x) = g_{E}(x) = x$. Now, the SQL query can be expressed in our formalism as follows:
\begin{align*}
  \VIEW[A,C]{Q} = \VSUM_{B}\VSUM_{D}\VSUM_{E} \VIEW[A,B]{R} \VPROD \VIEW[A,C,E]{S} \VPROD \VIEW[C,D]{T}
\end{align*}

The computation of the aggregate \texttt{SUM(R.B$\,$*$\,$T.D$\,$*$\,$S.E)} now happens over payloads. 
\punto
\end{example}

One key benefit of using relations over rings is avoiding the intricacies of incremental computation under classical multiset semantics caused by non-commutativity of inserts and deletes. Our data model simplifies delta processing by representing both inserts and deletes as tuples, with the distinction that they map to positive and respectively negative ring values. This uniform treatment allows for simple delta rules for the three operators of our query language.

\bigskip

\nop{
This sum-product formulation covers many computational problems such as conjunctive query evaluation, inference in probabilistic graphical models, and matrix chain multiplication~\cite{FAQ:PODS:2016}. 
}


\section{Factorized Ring Computation}
\label{sec:factorized_ring_computation}

In this section, we introduce a static query evaluation framework based on factorized computation and data rings. In the next section, we adapt it to incremental maintenance.


\begin{figure*}[t]
\hspace*{-4mm}
\scalebox{0.8}{
\subfloat[Variable order $\omega$]
{
  \label{fig:example_variableorder}
  \begin{minipage}[b]{2.4cm}
    \centering
    \hspace*{-3mm}
    \begin{small}
      \begin{tikzpicture}[xscale=0.45, yscale=0.43]

        \draw[rotate=73,line width=0.1mm,fill opacity=0.8,fill=magenta!40] (-2.6,-0.2) ellipse (2.1cm and 0.5cm);

        \draw[rotate=-72.2,line width=0.1mm,fill opacity=0.8,fill=green!60] (4.15,-0.2) ellipse (3.65cm and 0.6cm);

        \draw[rotate=71,line width=0.1mm,fill opacity=0.8,fill=blue!40] (-5,-2.3) ellipse (2.1cm and 0.5cm);

        \node at (0, -1) (A) {\phantom{\{}A\phantom{\}}};
        \node at (-1, -4) (B) {\phantom{\{}B\phantom{\}}} edge[-, shorten >=0.1cm, shorten <=0.1cm] (A);
        \node at (1, -4) (C) {\phantom{\{}C\phantom{\}}} edge[-, shorten >=0.1cm, shorten <=0.1cm] (A);
        \node at (0, -7) (D) {\phantom{\{}D\phantom{\}}} edge[-, shorten >=0.1cm, shorten <=0.1cm] (C);
        \node at (2, -7) (E) {\phantom{\{}E\phantom{\}}} edge[-, shorten >=0.1cm, shorten <=0.1cm] (C);
        
        \node at (-1.1, -5) {\normalsize $\VIEW{R}$};
        \node at (-0.1, -8) {\normalsize $\VIEW{T}$};
        \node at (2.25, -8) {\normalsize $\VIEW{S}$};

      \end{tikzpicture}%
      \\[1ex]
      \scalebox{0.85} {
        \begin{tabular}{@{~~}l}
          $dep(A) = \emptyset$\\[1ex]
          $dep(B)=\{A\}$\\[1ex]
          $dep(C)=\{A\}$\\[1ex]
          $dep(D)=\{C\}$\\[1ex]
          $dep(E)=\{A,C\}$\\[8ex]
        \end{tabular}
      }
    \end{small}
    \vspace{-5.7ex}
  \end{minipage}
}
\quad
\subfloat[View tree over $\omega$ and $\mathcal{F} = \emptyset$]
{
  \label{fig:example_viewtree}
  \hspace*{-1mm}
  \begin{minipage}[b]{4cm}
    \small
    \hspace{3pt}
    \begin{tikzpicture}[xscale=0.7, yscale=0.24]

      \node at (0, -1.5) (A) {$\VIEW[~]{V^{@A}_{RST}}$};
      \node at (-1.5, -5) (B) {$\VIEW[A]{V^{@B}_{R}}$} edge[-] (A);
      \node at (1.5, -5) (C) {$\VIEW[A]{V^{@C}_{ST}}$} edge[-] (A);
      \node at (0.5, -9) (D) {$\VIEW[C]{V^{@D}_{T}}$} edge[-] (C);
      \node at (2.5, -9) (E) {$\VIEW[A,C]{V^{@E}_{S}}$} edge[-] (C);
      
      \node at (-1.5, -8) {$\VIEW[A,B]{R}$} edge[-] (B);
      \node at (2.5, -12) {$\VIEW[A,C,E]{S}$} edge[-] (E);
      \node at (0.5, -12) {$\VIEW[C,D]{T}$} edge[-] (D);

    \end{tikzpicture}
    \\[3pt]
    \hspace*{3pt}
    \scalebox{0.85}{\parbox{\linewidth}{%
      \begin{align*}
        \VIEW[~]{V^{@A}_{RST}} &= \VSUM_{A} \big(\VIEW[A]{V^{@B}_{R}} \VPROD \VIEW[A]{V^{@C}_{ST}}\big) \\[2pt]
        \VIEW[A]{V^{@B}_{R}} &= \VSUM_{B} \VIEW[A,B]{R} \\[2pt] 
        \VIEW[A]{V^{@C}_{ST}} &= \VSUM_{C} \big(\VIEW[C]{V^{@D}_{T}} \VPROD \VIEW[A,C]{V^{@E}_{S}}\big) \\[2pt]
        \VIEW[C]{V^{@D}_{T}} &= \VSUM_{D} \VIEW[C,D]{T}\\[2pt]
        \VIEW[A,C]{V^{@E}_{S}} &= \VSUM_{E} \VIEW[A,C,E]{S}
      \end{align*}
    }}
    \vspace{-3pt}
  \end{minipage}
}
\qquad
\subfloat[Database $\db$]
{
  \hspace*{3mm}
  \label{fig:example_payloads_database}
  \begin{minipage}[b]{2cm} 
    \scriptsize    
    \begin{tabular}{@{}l@{~~}l@{~$\to$~}l@{}}
      $A$ & $B$ & $\VIEW{R}[A,B]$\\\toprule
      $a_1$ & $b_1$ & $p_1$ \\
      $a_1$ & $b_2$ & $p_2$\\  
      $a_2$ & $b_3$ & $p_3$\\
      $a_3$ & $b_4$ & $p_4$\\\bottomrule
    \end{tabular}
    \\[6ex]
    \begin{tabular}{@{}l@{~~}l@{~~}l@{~$\to$~}l@{}}
      $A$ & $C$ & $E$ & $\VIEW{S}[A,C,E]$ \\\toprule
      $a_1$ & $c_1$ & $e_1$ & $p_5$\\
      $a_1$ & $c_1$ & $e_2$ & $p_6$\\
      $a_1$ & $c_2$ & $e_3$ & $p_7$\\
      $a_2$ & $c_2$ & $e_4$ & $p_8$\\\bottomrule
    \end{tabular}
    \\[6ex]
    \begin{tabular}{@{}l@{~~}l@{~$\to$~}l@{}}
      $C$ & $D$ & $\VIEW{T}[C,D]$ \\\toprule
      $c_1$ & $d_1$ & $p_9$\\
      $c_2$ & $d_2$ & $p_{10}$\\
      $c_2$ & $d_3$ & $p_{11}$\\
      $c_3$ & $d_4$ & $p_{12}$\\\bottomrule
    \end{tabular}
    \vspace{0.4em}

  \end{minipage}
}
\qquad
\subfloat[{\tt COUNT} query over $\mathcal{D}$]
{
  \label{fig:example_payloads_count}
  \scalebox{0.95}{
  \hspace{-0.2cm}
  \begin{minipage}[b]{2.8cm}
    \scriptsize
    \begin{tikzpicture}[xscale=0.75, yscale=0.35]
      \node [anchor=north west] at (-5, 9) {
        \begin{tabular}{@{}l@{\,}  @{\,}c@{\,}c@{\,}l@{}}
          & $()$ & $\rightarrow$ & \ $\VIEW[\;]{V^{@A}_{RST}}$ \\[1ex]\toprule
          & $()$ & $\rightarrow$ & 10 \\\bottomrule
        \end{tabular}
      };

      \node [anchor=north west] at (-6, 5.5) {
        \begin{tabular}{@{}l@{\,} @{\,}c@{\,}c@{\,}l@{}}
          & $\mathsf{A}$ & $\to$ & $\VIEW[A]{V^{@B}_{R}}$ \\[1ex]\toprule
          & $a_1$ & $\rightarrow$ & 2 \\
          & $a_2$ & $\rightarrow$ & 1 \\
          & $a_3$ & $\rightarrow$ & 1\\\bottomrule
        \end{tabular}
      };

      \node [anchor=north west] at (-4, 5.5) {
        \begin{tabular}{@{}l@{\,} @{\,}c@{\,}c@{\,}l@{}}
          & $\mathsf{A}$ & $\rightarrow$ & $\VIEW[A]{V^{@C}_{ST}}$ \\[1ex]\toprule
          & $a_1$ & $\rightarrow$ & 4 \\
          & $a_2$ & $\rightarrow$ & 2 \\\\\bottomrule 
        \end{tabular}
      };

      \node [anchor=north west] at (-5, 0.5) {
        \begin{tabular}{@{}l@{\,} @{\,}c@{\,}c@{\,}l@{}}
          & $\mathsf{C}$ & $\to$ & $\VIEW[C]{V^{@D}_{T}}$ \\[1ex]\toprule
          & $c_1$ & $\rightarrow$ & 1 \\
          & $c_2$ & $\rightarrow$ & 2 \\
          & $c_3$ & $\rightarrow$ & 1 \\\bottomrule
        \end{tabular}
      };

      \node [anchor=north west] at (-5, -4) {
        \begin{tabular}{@{}l@{\,} @{\,}c@{\,}c@{\,}c@{\,}l@{}}
          & $\mathsf{A}$ & $\mathsf{C}$ & $\to$ & $\VIEW[A,C]{V^{@E}_{S}}$ \\[1ex]\toprule
          & $a_1$ & $c_1$ & $\rightarrow$ & 2 \\
          & $a_1$ & $c_2$ & $\rightarrow$ & 1 \\
          & $a_2$ & $c_2$ & $\rightarrow$ & 1 \\\bottomrule
        \end{tabular}
      };

    \end{tikzpicture}
  \end{minipage}
  }
}
\quad
\subfloat[Conjunctive query over $\mathcal{D}$]
{
  \label{fig:example_payloads_conjunctive}
  \begin{minipage}[b]{6cm}
    \centering
    \begin{tikzpicture}[xscale=2.3, yscale=1.1]

      \node [text=blue, anchor=north west] at (2.55, 1) {
        \scriptsize
        \begin{tabular}{@{}l@{\,} @{\,}c@{\,}c@{\,}l@{}}
          & $()$ & $\to$ & $\VIEW[\;]{V^{@A}_{RST}}$ \\[1ex]\toprule 
           & $\tuple{}$ & $\rightarrow$ &
            \begin{tabular}{@{}l@{\,}!{\vrule width 0.03em}@{\,}c@{}c@{}c@{}}
              & $\mathsf{A}$ & & \\
              \specialrule{.03em}{0em}{0em} 
              & $a_1$ & $\rightarrow$ & $8$ \\
              & $a_2$ & $\rightarrow$ & $2$ \\
            \end{tabular}\\\bottomrule 
        \end{tabular}
      };

      \node [text=red, anchor=north east] at (4.6, 1) {
        \scriptsize
        \begin{tabular}{@{}l@{\,}  @{\,}c@{\,}c@{\,}l@{}}
          & $()$ & $\rightarrow$ & \ $\VIEW[\;]{V^{@A}_{RST}}$ \\[1ex]\toprule
            & $\tuple{}$ & $\rightarrow$ & 
            \begin{tabular}{@{}l@{\,}!{\vrule width 0.03em}@{\,}c@{\,}c@{\,}c@{\,}c@{}c@{}c@{}}
              & $\mathsf{A}$ & $\mathsf{B}$ & $\mathsf{C}$ & $\mathsf{D}$ & & \\
              \specialrule{.03em}{0em}{0em} 
              & $a_1$ & $b_1$ & $c_1$ & $d_1$ & $\rightarrow$ & $2$ \\
              & $a_1$ & $b_1$ & $c_2$ & $d_2$ & $\rightarrow$ & $1$ \\
              & $a_1$ & $b_1$ & $c_2$ & $d_3$ & $\rightarrow$ & $1$ \\
              & $a_1$ & $b_2$ & $c_1$ & $d_1$ & $\rightarrow$ & $2$ \\
              & $a_1$ & $b_2$ & $c_2$ & $d_2$ & $\rightarrow$ & $1$ \\
              & $a_1$ & $b_2$ & $c_2$ & $d_3$ & $\rightarrow$ & $1$ \\
              & $a_2$ & $b_3$ & $c_2$ & $d_2$ & $\rightarrow$ & $1$ \\
              & $a_2$ & $b_3$ & $c_2$ & $d_3$ & $\rightarrow$ & $1$ \\
            \end{tabular}\\\bottomrule
        \end{tabular}
      };

      \node [text=blue, anchor=north west] at (2.55, -0.78) {
        \scriptsize
        \begin{tabular}{@{}l@{\,} @{\,}c@{\,}c@{\,}l@{}}
          & $\mathsf{A}$ & $\rightarrow$ & $\VIEW[A]{V^{@C}_{ST}}$ \\[1ex]\toprule
           & $a_1$ & $\rightarrow$ &
            \begin{tabular}{@{}l@{\,}!{\vrule width 0.03em}@{\,}c@{\,}c@{\,}c@{}}
              & $\mathsf{C}$ & & \\
              \specialrule{.03em}{0em}{0em} 
              & $c_1$ & $\rightarrow$ & $2$ \\
              & $c_2$ & $\rightarrow$ & $2$ \\
            \end{tabular} \\
          \rule{0mm}{4mm} & $a_2$ & $\rightarrow$ &
            \begin{tabular}{@{}l@{\,}!{\vrule width 0.03em}@{\,}c@{\,}c@{\,}c@{}}
                & $\mathsf{C}$ & & \\
                \specialrule{.03em}{0em}{0em} 
                & $c_2$ & $\rightarrow$ & $2$ \\
            \end{tabular}\\\bottomrule 
        \end{tabular}
      };

      \node [text=red, anchor=south east] at (4.6, -5) {
        \scriptsize
        \begin{tabular}{@{}l@{\,} @{\,}c@{\,}c@{\,}l@{}}
          & $\mathsf{A}$ & $\to$ & \ $\VIEW[A]{V^{@C}_{ST}}$ \\[1ex]\toprule
           & $a_1$ & $\rightarrow$ & 
            \begin{tabular}{@{}l@{\,}!{\vrule width 0.03em}@{\,}c@{\,}c@{\,}c@{\,}c@{}}
              & $\mathsf{C}$ & $\mathsf{D}$ & & \\
              \specialrule{.03em}{0em}{0em} 
              & $c_1$ & $d_1$ & $\rightarrow$ & $2$ \\
              & $c_2$ & $d_2$ & $\rightarrow$ & $1$ \\
              & $c_2$ & $d_3$ & $\rightarrow$ & $1$ \\
            \end{tabular}\\
          \rule{0mm}{6mm} & $a_2$ & $\rightarrow$ & 
            \begin{tabular}{@{}l@{\,}!{\vrule width 0.03em}@{\,}c@{\,}c@{\,}c@{\,}c@{}}
                & $\mathsf{C}$ & $\mathsf{D}$ & & \\
                \specialrule{.03em}{0em}{0em} 
                & $c_2$ & $d_2$ & $\rightarrow$ & $1$ \\
                & $c_2$ & $d_3$ & $\rightarrow$ & $1$ \\
            \end{tabular}\\\bottomrule
        \end{tabular}
      };

      \node [anchor=south west] at (2.5, -5) {
        \scriptsize
        \begin{tabular}{@{}l@{\,} @{\,}c@{\,}c@{\,}c@{\,}c@{}}
          & $\mathsf{A}$ & $\mathsf{C}$ & $\to$ & $\VIEW[A,C]{V^{@E}_{S}}$ \\[1ex]\toprule
           & $a_1$ & $c_1$ & $\rightarrow$ & 
            \begin{tabular}{@{}l@{\,}!{\vrule width 0.03em}@{\,}c@{\,}c@{\,}c@{}}
              & & & \\[-2ex]
              \specialrule{.03em}{0em}{0em} 
              & $\tuple{}$ & $\rightarrow$ & $2$ \\
            \end{tabular} \\
          \rule{0mm}{3mm} & $a_1$ & $c_2$ & $\rightarrow$ & 
            \begin{tabular}{@{}l@{\,}!{\vrule width 0.03em}@{\,}c@{\,}c@{\,}c@{}}
                & & & \\[-2ex]
                \specialrule{.03em}{0em}{0em} 
                & $\tuple{}$ & $\rightarrow$ & $1$ \\
            \end{tabular}\\
          \rule{0mm}{3mm} & $a_2$ & $c_2$ & $\rightarrow$ &
            \begin{tabular}{@{}l@{\,}!{\vrule width 0.03em}@{\,}c@{\,}c@{\,}c@{}}
                & & & \\[-2ex]
                \specialrule{.03em}{0em}{0em} 
                & $\tuple{}$ & $\rightarrow$ & $1$ \\
            \end{tabular}\\\bottomrule
        \end{tabular}
      };

      \node [anchor=north west] at (1.65, 1) {
        \scriptsize
        \begin{tabular}{@{}l@{\,} @{\,}c@{\,}c@{\,}c@{}}
          & $\mathsf{A}$ & $\to$ & $\VIEW[A]{V^{@B}_{R}}$ \\[1ex]\toprule
           & $a_1$ & $\rightarrow$ &
            \begin{tabular}{@{}l@{\,}!{\vrule width 0.03em}@{\,}c@{\,}c@{\,}c@{}}
              & $\mathsf{B}$ & & \\
              \specialrule{.03em}{0em}{0em} 
              & $b_1$ & $\rightarrow$ & $1$ \\
              & $b_2$ & $\rightarrow$ & $1$ \\
            \end{tabular} \\
          \rule{0mm}{4mm} & $a_2$ & $\rightarrow$ &
            \begin{tabular}{@{}l@{\,}!{\vrule width 0.03em}@{\,}c@{\,}c@{\,}c@{}}
                & $\mathsf{B}$ & & \\
                \specialrule{.03em}{0em}{0em} 
                & $b_3$ & $\rightarrow$ & $1$ \\
            \end{tabular} \\
          \rule{0mm}{4mm} & $a_3$ & $\rightarrow$ &
            \begin{tabular}{@{}l@{\,}!{\vrule width 0.03em}@{\,}c@{\,}c@{\,}c@{}}
                & $\mathsf{B}$ & & \\
                \specialrule{.03em}{0em}{0em} 
                & $b_4$ & $\rightarrow$ & $1$ \\
            \end{tabular}\\\bottomrule 
        \end{tabular}
      };

      \node [anchor=south west] at (1.65, -5) {
        \scriptsize
        \begin{tabular}{@{}l@{\,} @{\,}c@{\,}c@{\,}c@{}}
          & $\mathsf{C}$ & $\to$ & $\VIEW[C]{V^{@D}_{T}}$ \\[1ex]\toprule
           & $c_1$ & $\rightarrow$ &
            \begin{tabular}{@{}l@{\,}!{\vrule width 0.03em}@{\,}c@{\,}c@{\,}c@{}}
                & $\mathsf{D}$ & & \\
                \specialrule{.03em}{0em}{0em} 
                & $d_1$ & $\rightarrow$ & $1$ \\
            \end{tabular} \\
          \rule{0mm}{6mm} & $c_2$ & $\rightarrow$ & 
            \begin{tabular}{@{}l@{\,}!{\vrule width 0.03em}@{\,}c@{\,}c@{\,}c@{}}
              & $\mathsf{D}$ & & \\
              \specialrule{.03em}{0em}{0em} 
              & $d_2$ & $\rightarrow$ & $1$ \\
              & $d_3$ & $\rightarrow$ & $1$ \\
            \end{tabular} \\
          \rule{0mm}{4.5mm} & $c_3$ & $\rightarrow$ &
            \begin{tabular}{@{}l@{\,}!{\vrule width 0.03em}@{\,}c@{\,}c@{\,}c@{}}
                & $\mathsf{D}$ & & \\
                \specialrule{.03em}{0em}{0em} 
                & $d_4$ & $\rightarrow$ & $1$ \\
            \end{tabular}\\\bottomrule 
        \end{tabular}
      };
    \end{tikzpicture}
  \end{minipage}
}
}
\caption{
(a) Variable order $\omega$ of the natural join of {\color{magenta!80}$\VIEW{R}$}, {\color{goodgreen}$\VIEW{S}$}, {\color{blue!80}$\VIEW{T}$}.
(b) View tree $\tau$ over $\omega$ and without free variables. 
(c) Database $\db$ over a ring $\RING$, where $\{p_i\}_{i\in[12]}\subseteq\RING$. 
(d) Computing {\tt COUNT} using $\tau$ and the $\mathbb{Z}$ ring, where $\forall i\in[12]: p_i=1$. 
(e) Computing the query from Example~\ref{ex:relational_ring} using $\tau$ and the relational ring, where $\forall i\in[12]: p_i=\{ () \to 1 \}$. The red views (rightmost column) have payloads storing the listing representation of the intermediate and final query results. The blue views (middle) encode a factorized representation of these results distributed over their payloads. The black views are the same for both representations.
}
\label{fig:example_payloads}
\end{figure*}
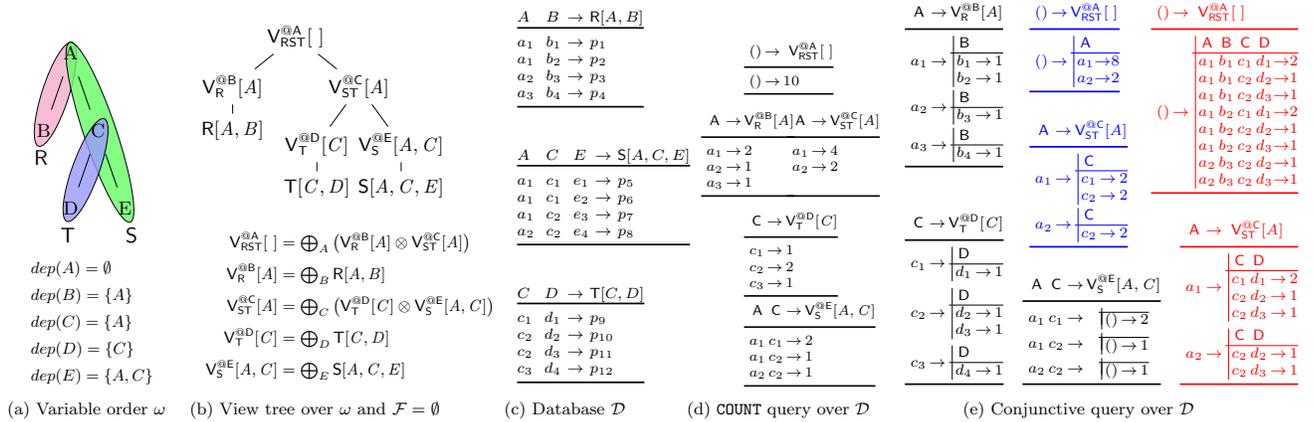

\begin{figure}[t]
\centering
\setlength{\tabcolsep}{3pt}
\begin{tabular}{@{}c@{}c@{~~~}l}
  \toprule
  \multicolumn{3}{c}{$\tau$ (\text{variable order} $\omega$, \text{free variables} $\mathcal{F}$)} \\
  \midrule
  \multicolumn{3}{l}{\MATCH $\omega$:} \\
  \midrule 
  \phantom{a} & $\VIEW{R}$\hspace*{2.5em} & $\VIEW[\mathit{\sch(\VIEW{R})}]{R}$ \\
  \cmidrule{2-3} \\[-6pt] 
  &
  \begin{minipage}[b]{3cm}
    \begin{tikzpicture}[xscale=0.48, yscale=1]
      \node at (0,-2)  (n4) {$X$};
      \node at (-1,-3)  (n1) {$\omega_1$} edge[-] (n4);
      \node at (0,-3)  (n2) {$\ldots$};
      \node at (1,-3)  (n3) {$\omega_k$} edge[-] (n4);
      \node at (0,-4.5) {~};
    \end{tikzpicture}
    \vspace{2.85cm}
  \end{minipage}
  &
  \begin{minipage}[b]{8cm}
    \hspace{0.6cm}
    \begin{tikzpicture}[xscale=0.9, yscale=1]
      \node at (0,-2)  (n4) {$\VIEW[\mathit{keys}]{V^{@X}_{rels}}$};
      \node at (-1,-3)  (n1) {$\tau(\omega_1, \mathcal{F})$} edge[-] (n4);
      \node at (0,-3)  (n2) {$\ldots$};
      \node at (1,-3)  (n3) {$\tau(\omega_k, \mathcal{F})$} edge[-] (n4);
      \node at (2.5,-2.5) {$\text{, where}$};  
    \end{tikzpicture} \\[1ex]
    $\LET$\\[0.5ex]
    $\TAB\VIEW[\mathit{keys_i}]{V^{@\omega_i}_{rels_i}} = \text{ root of } \tau(\omega_i, \mathcal{F}), \forall i\in[k], $\\[0.5ex]
    $\TAB\mathit{keys}=\mathit{dep}(X) \cup (\mathcal{F} \cap \bigcup_{i\in[k]}\mathit{keys_i}),$\\[0.5ex]
    $\TAB\mathsf{rels}=\bigcup_{i\in[k]}\mathsf{rels}_i,$\\[0.5ex]
    $\TAB\mathit{V}[{\it keys}]=\VPRODBIG_{i \in [k]} \VIEW[\mathit{keys_i}]{V^{@\omega_i}_{rels_i}},$\\[0.5ex]
    $\IN$\\[0.5ex]    
    $\TAB\VIEW[\mathit{keys}]{V^{@X}_{rels}} =\left\{
      \begin{array}{l}
        V[{\it keys}] \hspace*{2.2em}\text{,\; if } X \in \mathcal{F}\\
        \VSUM_{X} \mathit{V}[{\it keys}] \text{,\; otherwise.}
      \end{array}
      \right.$\\
  \end{minipage}
  \\
  \bottomrule
\end{tabular}
\caption{Creating a view tree $\tau(\omega, \mathcal{F})$ for a variable order $\omega$ and a set of free variables $\mathcal{F}$.}
\label{fig:static_view_tree_algo}
\end{figure}
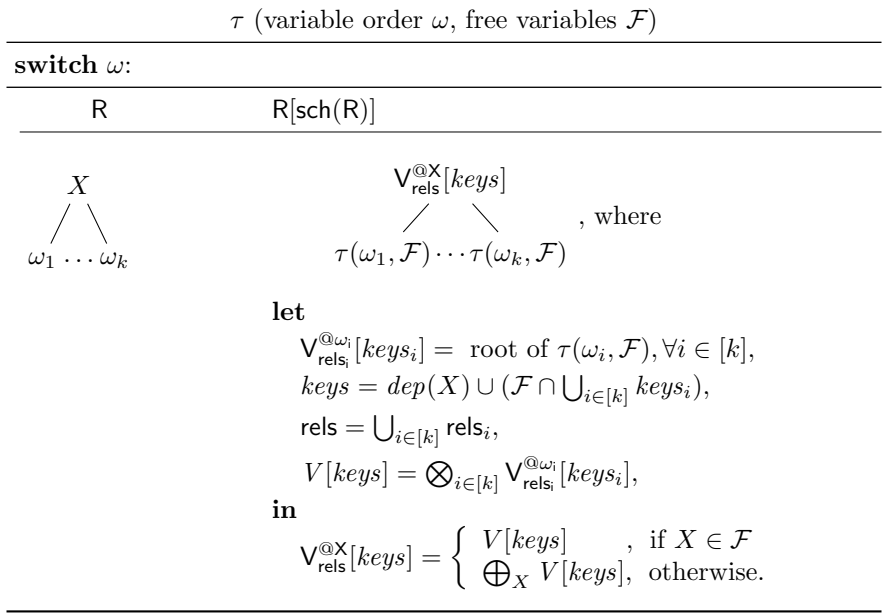


{\bf Variable Orders.} Classical query evaluation uses query plans that dictate the order in which the relations are joined. We use slightly different plans, which we call variable orders, that dictate the  order in which we solve each join variable. They  may require to join several relations at the same time if these relations have the same variable. Our choice is motivated by the complexity of join evaluation: standard (relation-at-a-time) query plans are provably suboptimal, whereas variable orders can be optimal~\cite{Ngo:SIGREC:2013}.

Given a join query $Q$, a variable $X$ {\em depends} on a variable $Y$ if both are in the schema of a relation in $Q$.

\begin{definition}[adapted from \cite{Olteanu:FactBounds:2015:TODS}]\em\label{def:vo}
A {\em variable order} $\omega$ for a join query $Q$ is a pair $(F,dep)$, where $F$ is a rooted forest with one node per variable in $Q$, and {\em dep} is a function mapping each variable $X$ to a set of variables in $F$. It satisfies the following constraints:
\begin{itemize}
\item For each relation in $Q$, its variables lie along the same root-to-leaf path in $F$.
\item For each variable $X$, $dep(X)$ is the subset of its ancestors in $F$ on which the variables in the subtree rooted at $X$ depend.
\end{itemize}
\end{definition}

Figure~\ref{fig:example_variableorder} gives a variable order for the natural join of three relations. Variable $D$ has ancestors $A$ and $C$, yet it only depends on $C$ since $C$ and $D$ appear in the same relation $T$ and $D$ does not occur in any relation together with $A$. Thus, $dep(D)=\{C\}$. Given $C$, the variables $D$ and $E$ are independent of each other. In case $Q$ has free variables, then we prefer variable orders for $Q$ that have all free variables on top of the bound variables~\cite{BKOZ:PVLDB:2013,Olteanu:FactBounds:2015:TODS}.

\nop{
\milos{the paragraph below talks about factorized representations while here we talk about factorized computation}
Although lossless, factorized representations of the query result can be arbitrarily smaller than the standard listing representation. Whereas a listing representation of the query result would contain the Cartesian product of sets of values for $D$ and $E$, a factorized representation would avoid the explicit materialization of this product. Aggregates defined by ring expressions over the query variables can be computed in one pass over factorized joins whose variable orders have the free variables on top of the bound ones~\cite{BKOZ:PVLDB:2013}. 
}


{\bf View Trees.} Our framework relies on a variable order $\omega$ for the input query $Q$ to describe the structure of computation and indicate which variable marginalizations are pushed past joins. At each variable in $\omega$ we define a view that is a query over its children. The view at the root variable corresponds to the entire query $Q$. The tree of these views is called a {\em view tree}, which plays the role of a query plan in our framework. 
 
Figure~\ref{fig:static_view_tree_algo} gives an algorithm that constructs a view tree $\tau$ for a variable order $\omega$ of the input query $Q$ and the set of free variables $\mathcal{F}$ of $Q$. The input variable order is extended with relations at leaves under their lowest variable (as children or further below). The views in the view tree are defined over the input relations or the views at children. We use the notation $\VIEW[keys]{V^{@X}_{rels}}$ to state that the view $\VIEW{V}$ is recursively defined over the input relations {\sf rels}, has free variables $keys$, and is at the variable $X$ in $\omega$; in case of a view for an input relation $\VIEW{R}$, we use the simplified notation $\VIEW[\sch(\VIEW{R})]{R}$.

The base case (leaf in the extended variable order) is that of an input relation: We construct a view that is the relation itself. At a variable $X$ (inner node), we distinguish two cases: If $X$ is a bound variable, we construct a view that marginalizes out $X$ in the natural join of the child views; we thus first join on $X$, then apply the lifting function for $X$ on its values, and aggregate $X$ away. If $X$ is a free variable, however, we retain it in the view schema without applying the lifting function to its values. The schema of the view consists of those ancestor variables of $X$ in $\omega$ on which it depends and the free variables of its children.

\nop{
\milos{Restricting variable orders as below is only necessary for factorized payloads. In other cases, our framework works with any variable order.}
The algorithm works on {\em any} variable order that is valid for the input query. The variable orders valid for queries with free (group-by) variables~\cite{BKOZ:PVLDB:2013,FAQ:PODS:2016} have the free variables placed above the bound variables. This leads to view trees where marginalization of bound variables happens as early as possible in a bottom-up evaluation.
}

\begin{example}\em
Figure~\ref{fig:example_payloads} gives the view tree constructed by our algorithm for the given variable order and the empty set of free variables. The figure also shows the contents of the views computing the {\tt COUNT} query over the database $\mathcal{D}$. 
\punto
\end{example}

By default, the algorithm constructs one view per variable in the variable order $\omega$. A wide relation (with many variables) leads to long branches in $\omega$ with variables that are only local to this relation. This is, for instance, the case of our retailer dataset used in Section~\ref{sec:experiments}. Such long branches create long chains of views, where each view marginalizes one bound variable over its child view in the chain. For practical reasons, we compose such long chains into a single view that marginalizes several variables at a time. 

\bigskip

\nop{
\begin{example}\em
Figure~\ref{fig:example_payloads} gives the view tree constructed by our algorithm for the given variable order and the empty set of free variables. The figure also shows the contents of the views computing the {\tt COUNT} query over the database $\mathcal{D}$. 

Consider now an extension of our running example where the relation $\VIEW{S}[A,C,E,F]$ has one more bound variable $F$ placed under $E$ in the variable order. The views for $F$ and $E$ are defined by:
{\color{blue}
\begin{align*}
 \VIEW[A,C,E]{V^{@F}_{S}} &= \VSUM_{F}{\VIEW[A,C,E,F]{S}} \\
 \VIEW[A,C]{V^{@E}_{S}} &= \VSUM_{E}{\VIEW[A,C,E]{V^{@F}_{S}}}
\end{align*}
}
We can compose the two views into one equivalent view:
{\color{blue}
\begin{align*}
\VIEW[A,C]{V^{@E}_{S}} &= \VSUM_{E,F} \VIEW[A,C,E,F]{S} = \VSUM_{E} (\VSUM_{F} \VIEW[A,C,E,F]{S}).
\end{align*}
where the lifting function $g_{E,F}(e,f) = g_{E}(e) * g_{F}(f).$
\punto
}
\end{example}
}

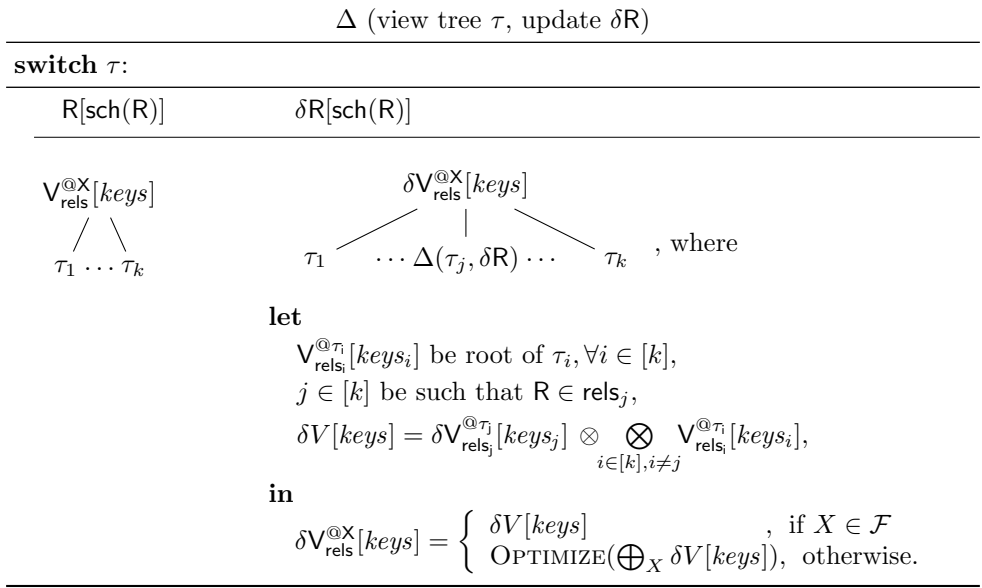
\begin{figure}[t]
\centering
\setlength{\tabcolsep}{3pt}
\begin{tabular}{@{}c@{}c@{~~~~}l}
    \toprule
    \multicolumn{3}{c}{$\Delta$ (\text{view tree} $\tau$, \text{update} $\VIEW{\delta{R}}$)}\\
    \midrule
    \multicolumn{3}{l}{\MATCH $\tau$:}\\
    \midrule
    \phantom{ab} & $\VIEW{R}[\mathit{\sch(\VIEW{R})}]$\hspace*{2.5em} & $\VIEW{\delta{R}}[\mathit{\sch(\VIEW{R})}]$\\[2pt]
    \cmidrule{2-3} \\[-6pt]
    &    
    \begin{minipage}[t]{3cm}
        \vspace{-1.5cm}
        \begin{tikzpicture}[xscale=0.45, yscale=1]
            \node at (0,-2)  (n4) {$\VIEW[{\it keys}]{V^{@X}_{rels}}$};
            \node at (-1,-3)  (n1) {$\tau_1$} edge[-] (n4);
            \node at (0,-3)  (n2) {$\ldots$};
            \node at (1,-3)  (n3) {$\tau_k$} edge[-] (n4);
        \end{tikzpicture}        
    \end{minipage}
    &
    \begin{minipage}[t]{9cm}
        \begin{tikzpicture}[xscale=1.0, yscale=1]
            \node at (0,-2)  (n4) {$\delta\VIEW[{\it keys}]{V^{@X}_{rels}}$};
            \node at (-2,-3)  (n1) {$\tau_1$} edge[-] (n4);
            \node at (-1,-3)  (n2) {$\ldots$};
            \node at (0,-3)  (n1) {$\Delta(\tau_j, \VIEW{\delta{R}})$} edge[-] (n4);
            \node at (1,-3)  (n2) {$\ldots$};
            \node at (2,-3)  (n3) {$\tau_k$} edge[-] (n4);
            \node at (3,-2.8) {$\text{ ,  where}$};  
        \end{tikzpicture} \\[1ex]
    \hspace*{-1em}$\LET$\\[0.5ex]
    \hspace*{-1em}$\TAB\VIEW[{\it keys_i}]{V^{@\tau_i}_{rels_i}}\text{ be root of } \tau_i, \forall i\in[k], $\\[0.5ex]
    \hspace*{-1em}$\TAB j\in[k] \text{ be such that } \VIEW{R}\in\mathsf{rels}_j,$\\[0.5ex]
    \hspace*{-1em}$\TAB\delta{V}[{\it keys}]  = \delta\VIEW[{\it keys_j}]{V_{rels_j}^{@\tau_j}} \,\VPROD\!\! \underset{i\in[k],i\neq j}{\VPRODBIG}\!\VIEW[{\it keys_i}]{V^{@\tau_i}_{rels_i}},$\\[0.5ex]
        \hspace*{-1em}$\IN$\\[0.5ex]    
    $ \delta\VIEW[\mathit{keys}]{V^{@X}_{rels}}=\left\{
      \begin{array}{l}
        \delta{V}[{\it keys}] \hspace*{6.7em}\text{,\; if } X \in \mathcal{F}\\
        \textsc{Optimize}(\VSUM_{X} \delta{V}[{\it keys}]) \text{,\;  otherwise.}
      \end{array}
      \right.$\\
    \end{minipage}
    \\
    \bottomrule
\end{tabular}
\caption{Creating a delta view tree $\Delta(\tau,\VIEW{\delta{R}})$ for a view tree $\tau$ to accommodate an update $\VIEW{\delta{R}}$ to relation $\VIEW{R}$. }
\label{fig:dynamic_view_tree_algo}
\vspace*{1em}
\end{figure}

\section{Factorized Higher-Order IVM}
\label{sec:factorized_IVM}

We introduce incremental view maintenance in our factorized ring computation framework. In contrast to evaluation, the incremental maintenance of the query result may require the materialization and maintenance of views. An update to a relation $\VIEW{R}$ triggers changes in all views from the leaf $\VIEW{R}$ to the root of the view tree. 

\medskip

{\bf Updates.} The insertion (deletion) of a tuple $\textvec{t}$ into (from) a relation $\VIEW{R}$ is expressed as a delta relation $\delta\VIEW{R}$ that maps $\textvec{t}$ to $\RINGONE$ (and respectively $-\RINGONE$). In general, $\delta\VIEW{R}$ can be a relation, thus a collection of tuples mapped to payloads. The updated relation is then the union of the old relation and the delta relation: $\VIEW{R} := \VIEW{R} \VPLUS \delta\VIEW{R}$. 

\medskip

{\bf Delta Views.} For each view $\VIEW{V}$ affected by an update, a {\em delta view} $\VIEW{\delta{V}}$ defines the change in the view contents. In case the view $\VIEW{V}$ represents a relation $\VIEW{R}$, then $\VIEW{\delta{V}}=\VIEW{\delta{R}}$ if there are updates to $\VIEW{R}$ and  $\VIEW{\delta{V}}=\emptyset$ otherwise. If the view is defined using operators on other views, $\delta\VIEW{V}$ is derived using the following delta rules:
\begin{align*}
  \delta{(\VIEW{V_1} \VPLUS \VIEW{V_2})} &= \delta{\VIEW{V_1}} \VPLUS \delta{\VIEW{V_2}} \\
  \nop{\delta{(-\VIEW{V_1})} &= -\delta{\VIEW{V_1}} \\}
  \delta{(\VIEW{V_1} \VPROD \VIEW{V_2})} &= (\delta{\VIEW{V_1}} \VPROD \VIEW{V_2}) \VPLUS (\VIEW{V_1} \VPROD\delta{\VIEW{V_2}}) \VPLUS (\delta{\VIEW{V_1}} \VPROD \delta{\VIEW{V_2}})\\
  \delta{(\VSUM_{X}\VIEW{V})} &= \VSUM_{X}\delta{\VIEW{V}}
\end{align*}

The correctness of the rules follows from the associativity of $\VPLUS$ and the distributivity of $\VPROD$ over $\VPLUS$; $\VSUM_{X}$ is equivalent to the repeated application of $\VPLUS$ for the possible values of $X$. The derived delta views are subject to standard simplifications: If $\VIEW{V}$ is not defined over the updated relation $\VIEW{R}$, then its delta view $\VIEW{\delta{V}}$ is empty, and then we propagate this information using the identities $\emptyset \VPLUS \VIEW{V} = \VIEW{V} \VPLUS \emptyset = \VIEW{V}$ and $\emptyset \VPROD \VIEW{V} = \VIEW{V} \VPROD \emptyset = \emptyset$.

\medskip

{\bf Delta Trees.} Under updates to one relation, a view tree becomes a delta tree where the affected views become delta views. The algorithm in Figure~\ref{fig:dynamic_view_tree_algo} traverses the view tree $\tau$ top-down along the path from the root to the updated relation and replaces the views on that path with delta views. The {\sc Optimize} method rewrites delta view expressions to exploit factorized updates by avoiding the materialization of Cartesian products and pushing marginalization past joins; we explain this optimization in Section~\ref{sec:factorizable_updates}.

\begin{example}\em\label{ex:delta_view_tree}
Consider the query from Example~\ref{ex:sql_count} and its view tree from Figure~\ref{fig:example_viewtree}. The update $\VIEW{\delta{T}}$ triggers delta computation at each view from the leaf $\VIEW{T}$ to the root of the view tree:
\begin{align*}
  \delta\VIEW[C]{V^{@D}_{T}} &= \VSUM_{D} \delta\VIEW[C,D]{T} \\
  \delta\VIEW[A]{V^{@C}_{ST}} &= \VSUM_{C} \delta\VIEW[C]{V^{@D}_{T}} \VPROD \VIEW[A,C]{V^{@E}_{S}} \\
  \delta\VIEW[~]{V^{@A}_{RST}} &= \VSUM_{A} \VIEW[A]{V^{@B}_{R}} \VPROD \delta\VIEW[A]{V^{@C}_{ST}}
\end{align*}

Let us consider the ring $\mathbb{Z}$ and the lifting functions that map all values to $1$, and let $\VIEW{\delta{T}}[C,D]=\{ \tuple{c_1,d_1} \to -1, \tuple{c_2, d_2} \to 3 \}$. Given the contents of $\VIEW{V^{@E}_{S}}$ and $\VIEW{V^{@B}_{R}}$ from Figure~\ref{fig:example_payloads}, we now have:
 
\begin{center}
\begin{tabular}{@{}ccc@{}}
  \begin{tabular}[t]{@{}l@{~$\to$~}l}
  C & $\delta\VIEW{V^{@D}_{T}}[C]$  \\\toprule
  $c_1$ & $-1$ \\
  $c_2$ & $3$
  \end{tabular}
  &
  \begin{tabular}[t]{@{}l@{~$\to$~}l}
  A & $\delta\VIEW{V^{@C}_{ST}}[A]$\\\toprule
  $a_1$ & $1$ \\
  $a_2$ & $3$
  \end{tabular}
  &
  \begin{tabular}[t]{@{}l@{~$\to$~}l}
  $()$ & $\delta\VIEW{V^{@A}_{RST}}$ \\\toprule
  $\tuple{}$ & $5$
  \end{tabular}
\end{tabular}
\end{center}

A single-tuple update to $\VIEW{T}$ fixes the values for $C$ and $D$ and computing $\delta\VIEW{V^{@D}_{T}}$ takes constant time. The delta view $\delta\VIEW{V^{@C}_{ST}}$ iterates over all possible $A$-values for a fixed $C$-value, which takes linear time; $\delta\VIEW{V^{@A}_{RST}}$ incurs the same linear-time cost. A single-tuple update to either $\VIEW{R}$ or $\VIEW{S}$, however, fixes all variables on a leaf-to-root path in the delta view tree, giving a constant view maintenance cost.
\punto
\end{example}

In contrast to classical (first-order) IVM that only requires maintenance of the query result~\cite{Chirkova:Views:2012:FTD}, our approach is higher-order IVM as updates may trigger maintenance of several interrelated views. The fully-recursive IVM scheme of DBToaster~\cite{Koch:Ring:2010:PODS,DBT:VLDBJ:2014} creates one materialization hierarchy per relation in the query, whereas we use one view tree for all relations. This view tree relies on variable orders to decompose the query into views and factorize its computation and maintenance.

\medskip

{\bf Which Views to Materialize and Maintain?}
The answer to this question depends on which relations may change. 
The set of the updatable relations determines the possible delta propagation paths in a view tree, and these paths may use materialized views.

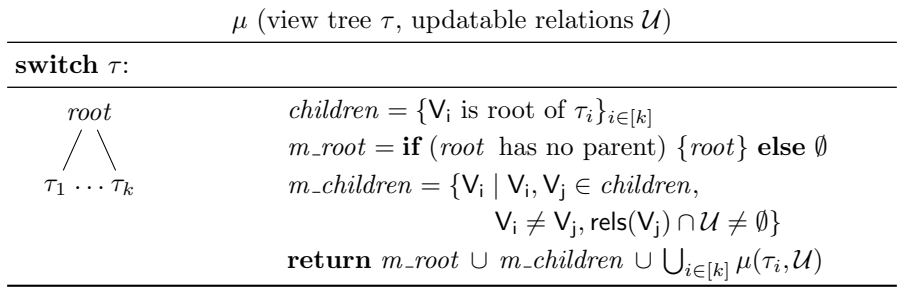
\begin{figure}[t]
\centering
\setlength{\tabcolsep}{3pt}
\begin{tabular}[t]{@{}c@{}c@{~~~}l}
    \toprule
    \multicolumn{3}{c}{$\mu$ (\text{view tree} $\tau$, \text{updatable relations} $\mathcal{U}$)}\\
    \midrule
    \multicolumn{3}{l}{\MATCH $\tau$:}\\
    \midrule 
    \phantom{ab}
    &
    \begin{minipage}[t]{3cm}
        \begin{tikzpicture}[xscale=0.45, yscale=1]
            \node at (0,-2)  (n4) {$\mathit{root}$};
            \node at (-1,-3)  (n1) {$\tau_1$} edge[-] (n4);
            \node at (0,-3)  (n2) {$\ldots$};
            \node at (1,-3)  (n3) {$\tau_k$} edge[-] (n4);
        \end{tikzpicture}
    \end{minipage}
    &
    \begin{minipage}[t]{8cm}
        \vspace{-4em}
        $\mathit{children} = \{ \VIEW{V_i} \text{ is root of } \tau_i \}_{i\in[k]}$\\[0.5ex]
        $\mathit{m\_root} = \IF\SPACE (\mathit{root} \,\text{ has no parent})\SPACE \{ \VIEW{\mathit{root}} \} \SPACE\ELSE\SPACE \emptyset$\\[0.5ex]
        $\mathit{m\_children} = 
          \{ \VIEW{V_i} \mid \VIEW{V_i}, \VIEW{V_j} \in \mathit{children}, $\\[0.5ex]
        $\TAB\TAB\TAB\TAB\TAB\TAB\TAB\SPACE \VIEW{V_i} \neq \VIEW{V_j}, \mathsf{rels}(\VIEW{V_j}) \cap \mathcal{U} \neq \emptyset \}$\\[0.5ex]
        $\RETURN\SPACE \mathit{m\_root} \,\cup\, \mathit{m\_children} \,\cup\, \bigcup_{i\in[k]} \mu(\tau_i, \mathcal{U})$
    \end{minipage}\\
    \bottomrule
\end{tabular}
\caption{Deciding which views in a view tree $\tau$ to materialize in order to support updates to a set of relations $\mathcal{U}$. The notation $\mathsf{rels}(\VIEW{V_j})$ denotes the relations of view $\VIEW{V_j}$.}
\label{fig:materialization_view_tree_algo}
\end{figure}

Propagating changes along a leaf-to-root path is computationally most effective if each delta view joins with sibling views that are already materialized. 
Figure~\ref{fig:materialization_view_tree_algo} gives an algorithm that reflects this idea: Given a view tree $\tau$ and a set of updatable relations $\mathcal{U}$, the algorithm traverses the tree top-down to discover the views that need to be materialized. The root of the view tree $\tau$ is always stored as it represents the query result. Every other view $\VIEW{V_i}$ is stored only if there exists a sibling view $\VIEW{V_j}$ defined over an updatable relation. The algorithm returns the set of chosen views.

\nop{
Figure~\ref{fig:materialization_view_tree_algo} depicts an algorithm that reflects this idea: Given a view tree $\tau$ and a set of updatable relations $\mathcal{U}$, the algorithm builds a {\em materialization tree} $\mu(\tau, \mathcal{U})$ that has the same structure as $\tau$ and where each node is a Boolean value indicating whether the corresponding view from $\tau$ should be materialized. The root view is always stored as it represents the query result, that is, it has no parent: $\VIEW{par}=\mathit{null}$; every other view $\VIEW{V}$ is stored only if it is used to compute the delta of its parent for updates to a relation over which $\VIEW{V}$ is not defined, that is, there are updatable relations for the parent and not for $\VIEW{V}$ itself: $(\VIEW{rels(par)} \setminus \VIEW{rels}) \cap \mathcal{U} \neq \emptyset$. 
}

\begin{example}\em
We continue with our query from Example~\ref{ex:delta_view_tree}. For updates to $\VIEW{T}$ only, that is, $\mathcal{U} = \{ \VIEW{T} \}$, we store the root $\VIEW{V^{@A}_{RST}}$ and the views $\VIEW{V^{@E}_{S}}$ and $\VIEW{V^{@B}_{R}}$ used to compute the deltas $\VIEW{\delta{V^{@C}_{ST}}}$ and $\VIEW{\delta{V^{@A}_{RST}}}$. Only the root view is affected by these changes and maintained as: 
\begin{align*}
\VIEW[~]{V^{@A}_{RST}} = \VIEW[~]{V^{@A}_{RST}} \VPLUS \VIEW[~]{\delta{V^{@A}_{RST}}}
\end{align*}

It is not necessary to maintain other views. If we would like to also support updates to both $\VIEW{R}$ and $\VIEW{S}$, then we would also need to materialize $\VIEW{V^{@C}_{ST}}$ and $\VIEW{V^{@D}_{T}}$. 
If no updates are supported, then only the root view is stored. 
\punto
\end{example}

For queries with free variables, several views in their (delta) view trees may be identical: This can happen when all variables in their keys are free and thus cannot be marginalized. For instance, a variable order $\omega$ for the query from Example~\ref{ex:sql_sum_aggregate} may have the variables $A$ and $C$ above all other variables, in which case their corresponding views are the same in the view tree for $\omega$. We then store only the top view out of these identical views. 

\medskip

{\bf IVM Triggers.}
For each updatable relation $\VIEW{R}$, our framework constructs a trigger procedure that takes as input an update $\VIEW{\delta{R}}$ and implements the maintenance schema of the corresponding delta view tree. This procedure also maintains all materialized views needed for the given update workload. 

A bulk of updates to distinct relations is handled as a sequence of updates, one per relation. Update sequences can also happen when updating a relation $\VIEW{R}$ that occurs several times in the query. The instances representing the same relation are at different leaves in the delta tree and lead to changes along multiple leaf-to-root paths.

\section{Factorizable Updates}
\label{sec:factorizable_updates}

Our focus so far has been on supporting updates represented by delta relations. We next consider an alternative approach that decomposes a delta relation into a union of factorizable relations. The cumulative size of the decomposition relations can be much less than the size of the original delta relation. Also, the complexity of propagating a factorized update can be much lower than that of its unfactorized (listing) representation, since the factorization makes explicit the independence between query variables and enables optimizations of delta propagation such as pushing marginalization past joins. Besides the factorized view computation, this is the second instance where our IVM approach exploits factorization.

Factorizable updates arise in many domains such as linear algebra and machine learning. Section~\ref{sec:applications} demonstrates how our framework can be used for the incremental evaluation of matrix chain multiplication, recovering prior work on this~\cite{NEK:SIGMOD:2014}.  Matrix chain computation can be phrased in our language of joins and aggregates, where matrices are binary relations. Changes to one row/column in an input matrix may be expressed as a product of two vectors. In general, an arbitrary update matrix can be decomposed into a sum of rank-$1$ matrices, each of them  expressible as products of vectors, using low-rank tensor decomposition methods~\cite{TensorDecomp:2009,TensorDecomposition:2017}.

\begin{example}\em
Arbitrary relations can be decomposed into a union of factorizable relations. The relation $\VIEW[A,B]{R}=\{(a_i,b_j) \to 1\mid i\in[n],j\in[m]\}$ can be decomposed as $\VIEW[A]{R_1}\VPROD\VIEW[B]{R_2}$, where $\VIEW[A]{R_1}=\{(a_i) \to 1\mid i\in[n]\}$ and $\VIEW[B]{R_2}=\{(b_j) \to 1\mid j\in[m]\}$. We thus reduced a relation of size $nm$ to two relations of cumulative size $n+m$. If $\VIEW{R}$ were a delta relation, the delta views on top of it would now be expressed over $\VIEW[A]{R_1}\VPROD\VIEW[B]{R_2}$ and their computation can be factorized as done for queries in Section~\ref{sec:factorized_ring_computation}. Product decomposition of relations can be done in linearithmic time in both the number of variables and the size of the relation~\cite{WSD:2008}. 

Consider now the relation $\VIEW[A,B]{R'}=\VIEW[A,B]{R}\VPLUS\{(a_{n+1},b_j) \to 1\mid j\in[m-1]\}$ with $\VIEW{R}$ as above. We can decompose each of the two terms in $\VIEW{R'}$ similarly to $\VIEW{R}$, yielding an overall $n+2m$ values instead of $nm+m-1$. A different decomposition with $n+m+3$ values is given by a factorizable over-approximation of $\VIEW{R'}$ compensated by a small product with negative payload: $\{(a_i)\to 1\mid i\in[n+1]\}\VPROD\{(b_j)\to 1\mid j\in[m]\}\VPLUS\{(a_{n+1})\to 1\}\VPROD\{(b_m)\to -1\}$.\punto
\end{example}

The {\sc Optimize} method used in the delta view tree algorithm in Figure~\ref{fig:dynamic_view_tree_algo} exploits the distributivity of join $\VPROD$ over marginalization $\VSUM_{X}$ to push $\VSUM_{X}$ past $\VPROD$ and down to the views with variable $X$. This optimization is reminiscent of pushing aggregates past joins in databases and variable elimination in probabilistic graphical models~\cite{FAQ:PODS:2016}. In case the delta views express Cartesian products, then they are not materialized but instead kept factorized.

\begin{example}\em
\label{ex:factorized-update}
Consider our query from Example~\ref{ex:delta_view_tree} under updates to relation $S$. Using the delta view tree derived for updates to $S$, the top-level delta is computed as follows:
\begin{align*}
\VIEW[~]{\delta{V^{@A}_{RST}}} = \VSUM_{A} \VIEW[A]{V^{@B}_{R}} \VPROD 
\underbrace{\big( \VSUM_{C} \VIEW[C]{V^{@D}_{T}} \VPROD 
  \underbrace{\big( \VSUM_{E} \VIEW[A,C,E]{\delta{S}} \,\big)}_{\VIEW[A,C]{\delta{V^{@E}_{S}}}}\big)}_{\VIEW[A]{\delta{V^{@C}_{ST}}}}
\end{align*}
A single-tuple update $\VIEW{\delta{S}}$ binds variables $A$, $C$, and $E$, and computing $\VIEW{\delta{V^{@A}_{RST}}}$ requires $\bigO{1}$ lookups in $\VIEW{V^{@D}_{T}}$ and $\VIEW{V^{@B}_{R}}$. An arbitrary-sized update $\VIEW{\delta{S}}$ can then be processed in $\bigO{|\VIEW{\delta{S}}|}$ time.

Assume now that $\VIEW{\delta{S}}$ is factorizable as $\VIEW[A,C,E]{\delta{S}} = \VIEW[A]{\delta{S_{A}}} \VPROD \VIEW[C]{\delta{S_{C}}} \VPROD \VIEW[E]{\delta{S_{E}}}$. In the construction of the delta view tree, the {\sc Optimize} method exploits this factorization to push the marginalization past joins at each variable; for example, the delta at $E$ becomes:
\begin{align*}
\VIEW[A,C]{\delta{V^{@E}_{S}}} &= \VSUM_{E} \VIEW[A]{\delta{S_{A}}} \VPROD \VIEW[C]{\delta{S_{C}}} \VPROD \VIEW[E]{\delta{S_{E}}} \\
&= \VIEW[A]{\delta{S_{A}}} \VPROD \VIEW[C]{\delta{S_{C}}} \VPROD \VSUM_{E} \VIEW[E]{\delta{S_{E}}}
\end{align*}
We also transform the top-level delta into a product of three views:
\begin{align*}
\VIEW[~]{\delta{V^{@A}_{RST}}} = 
&\big( \VSUM_{A} \VIEW[A]{V^{@B}_{R}} \VPROD \VIEW[A]{\delta{S_{A}}} \big) \VPROD \\
&\big( \VSUM_{C} \VIEW[C]{V^{@D}_{T}} \VPROD \VIEW[C]{\delta{S_{C}}} \big) \VPROD 
 \big( \VSUM_{E} \VIEW[E]{\delta{S_{E}}} \big)
\end{align*}
Computing this delta takes time proportional to the sizes of the three views representing the update:
$\bigO{\min(|\VIEW{V^{@B}_{R}}|, {|\VIEW{\delta{S_{A}}}|}) + \min(|\VIEW{V^{@D}_{T}}|, |\VIEW{\delta{S_{C}}}|) + |\VIEW{\delta{S_{E}}}|}$.
\punto
\end{example}


\section{IVM Variant for Cyclic Queries}
\label{sec:cyclic_queries}

Our framework supports arbitrary conjunctive queries. Whereas for ($\alpha$-)acyclic join queries the size of each view is asymptotically upper bounded by the size of the factorized join, for a cyclic join views may be larger in size than the (factorized) join result. This increase in space may however enable faster view maintenance.

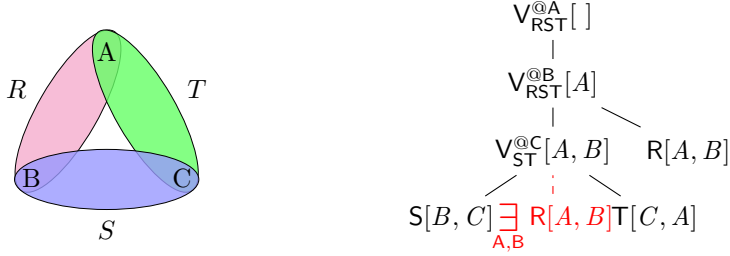
\begin{figure}\centering
  \begin{tabular}[c]{l@{~~~}l}  
    \begin{minipage}[b]{5cm}
      \begin{tikzpicture}[scale=0.5]
        \node at (-2.4, -2) (r) {$R$};
        \node at (2.4, -2) (r) {$T$};
        \node at (0, -5.75) (r) {$S$};
        \draw[rotate=60,line width=0.1mm,fill opacity=0.8,fill=magenta!40] (-2.75,-0.4) ellipse (2.45cm and 0.8cm);
       \draw[rotate=-60,line width=0.1mm,fill opacity=0.8,fill=green!60] (2.75,-0.4) ellipse (2.45cm and 0.8cm);
        \draw[rotate=0,line width=0.1mm,fill opacity=0.8,fill=blue!40] (0,-4.4) ellipse (2.45cm and 0.8cm);
        \node at (0, -1) (A) {A};
        \node at (-2, -4.35) (B) {B};
        \node at (2, -4.35) (C) {C};
      \end{tikzpicture}
      \vspace{2mm}
    \end{minipage}
    &
    \begin{minipage}[b]{7cm}
      \begin{tikzpicture}[xscale=0.9,yscale=0.9]
        \node at (0, 0) (A) {$\VIEW[~]{V^{@A}_{RST}}$};
        \node at (0, -1) (B) {$\VIEW[A]{V^{@B}_{RST}}$} edge[-] (A);
        \node at (0, -2) (C) {$\VIEW[A,B]{V^{@C}_{ST}}$} edge[-] (B);

        \node at (2, -2) (R) {$\VIEW[A,B]{R} \makebox[0pt][l]{$\phantom{\VIEW[A,B]{V^{@C}}}$} $} edge[-] (B);
        \node at (-1.5, -3) (S) {$\VIEW[B,C]{S}$} edge[-] (C);
        \node[color=red] at (-0, -3.15) (S) {$\VIEW[A,B]{\displaystyle\VEXISTS{A,B}{R}}$} edge[red,dashed] (C);
        \node at (1.5, -3) (T) {$\VIEW[C,A]{T}$} edge[-] (C);
      \end{tikzpicture}
    \end{minipage}
    \end{tabular}
\caption{(left) Hypergraph of the triangle query $Q_{\vartriangle}$; (right) View tree for the variable order $A - B - C$ with an indicator projection $\exists_{A,B}\VIEW{R}$.}
\label{fig:triangle_hypergraph_viewtree}
\end{figure}

\begin{example}\em\label{ex:triangle_query_ivm}
We consider the triangle query over the ring $\mathbb{Z}$:
\begin{align*}
\VIEW[~]{Q_{\vartriangle}} = \VSUM_{A}\VSUM_{B}\VSUM_{C} \VIEW[A,B]{R} \VPROD \VIEW[B,C]{S} \VPROD \VIEW[C,A]{T} 
\end{align*}

Figure~\ref{fig:triangle_hypergraph_viewtree} shows the hypergraph of $Q_{\vartriangle}$ and the view tree constructed for the variable order $A-B-C$ by placing each relation directly under its lowest variable. We assume all {relations} are of size $\bigO{N}$. Computing the triangle query from scratch using a worst-case optimal join algorithm takes $\bigO{N^{3/2}}$ time~\cite{Ngo:SIGREC:2013}.

In the given view tree (without the view in red), we first join $\VIEW{S}$ and $\VIEW{T}$ and then marginalize out $C$ in the join result. This view at node $C$ may contain $\bigO{N^2}$ pairs of $(A,B)$ values, which is larger than the worst-case size $\bigO{N^{3/2}}$. 
However, by materializing the view at $C$, we enable single-tuple updates to $R$ in constant time; single-tuple updates to other relations take $\bigO{N}$ time.

To avoid this large intermediate result, we can change the view tree by placing the relation $R$ under variable $C$. Then, joining all three relations at node $C$ takes $\bigO{N^{3/2}}$ time. Updates to any relation now cause recomputation of a $3$-way join, like in first-order IVM. For single-tuple updates, recomputing deltas takes $\bigO{N}$ as only two of the three variables are bound to constants. In contrast, the first approach trades off space for time: We need $\bigO{N^2}$ space but then support $\bigO{1}$ updates to one of the three relations.
\punto
\end{example}

The above example demonstrates how placing a relation under a different node in a view tree can create a cycle of relations and constrain the size of a view. This strategy, however, might not be always feasible or efficient: One relation might form multiple cycles of relations in different parts of a view tree -- for example, in a loop-$4$ cyclic query $Q_{\boxslash}$ with a chord,  the chord relation is part of two triangle subqueries. Since this relation cannot be duplicated in multiple subtrees (for correctness reasons so as to avoid multiplying the same payload several times instead of using it once), one would have to evaluate these subqueries in sequence, which yields a view tree that is higher and more expensive to maintain. 

\bigskip
{\bf Indicator Projections.}
Instead of moving relations in a view tree, we extend the tree with indicator projections that identify the active domains of these relations~\cite{FAQ:PODS:2016}. Such projections have no effect on the query result but can constrain view definitions (e.g., create cycles) and bring asymptotic savings in both space and time. 

We define a new unary operation $\VEXISTS{\mathcal{A}}{\VIEW{R}}$ that, given a relation $\VIEW{R}$ over schema $\mathcal{S}$ with payloads from a ring $(\RING, \RINGPLUS, \RINGPROD, \RINGZERO, \RINGONE)$, and a set of attributes $\mathcal{A} \subseteq \mathcal{S}$, projects tuples from $\VIEW{R}$ with non-$\RINGZERO$ payload on $\mathcal{A}$ and assigns to these tuples the payload $\RINGONE$. We define $\VEXISTS{\mathcal{A}}{\VIEW{R}}$ as:
\begin{align*}
  \forall\vecnormal{t} \in \Dom(\mathcal{A}):
  \left( \textstyle\VEXISTS{\mathcal{A}}{\VIEW{R}} \right)[\vecnormal{t}] = 
  \begin{cases} 
    \RINGONE & \exists\vecnormal{s} \in\Dom(\mathcal{S}), \vecnormal{s}\in \VIEW{R}, \vecnormal{t} = \pi_{\mathcal{A}}(\vecnormal{s})\\
    \RINGZERO & \text{otherwise}
  \end{cases}
\end{align*}

\nop{
The delta rule for $\VEXISTS{\mathcal{A}}$ is $\delta(\VEXISTS{\mathcal{A}}{\VIEW{R}}) = \VEXISTS{\mathcal{A}}(\VIEW{R} + \delta{\VIEW{R}}) - \VEXISTS{\mathcal{A}}{\VIEW{R}}$. This rule recomputes the query twice, once to insert new contents and once to delete the old contents, which clearly defeats the purpose of incremental computation. Note that some tuples might be unaffected by a given update $\delta{\VIEW{R}}$, so inserting those tuples and deleting them again is wasted work.

We observe that $\delta(\VEXISTS{\mathcal{A}}{\VIEW{R}})$ might change in the output only those tuples from $\VEXISTS{\mathcal{A}}{\delta{\VIEW{R}}}$. We exploit this restriction to refine the delta rule as: $\delta(\VEXISTS{\mathcal{A}}{\VIEW{R}}) = \VEXISTS{\mathcal{A}}{\delta{\VIEW{R}}} \VPROD (\VEXISTS{\mathcal{A}}(\VIEW{R} + \delta{\VIEW{R}}) - \VEXISTS{\mathcal{A}}{\VIEW{R}})$.
}

Indicator projections may change with updates to input relations. For instance, adding a tuple with a unique $\mathcal{A}$-value to $\VIEW{R}$ enlarges the result of $\VEXISTS{\mathcal{A}}\VIEW{R}$; similarly, deleting the last tuple with the given $\mathcal{A}$-value reduces the result. One change in the input may cause at most one change in the output: $|\delta{(\VEXISTS{\mathcal{A}}\VIEW{R})}| \le |\delta{\VIEW{R}}|$.

To facilitate the computation of $\delta{(\VEXISTS{\mathcal{A}}\VIEW{R})}$, we keep track of how many tuples with non-$\RINGZERO$ payloads project on each $\mathcal{A}$-value. For updating the payload of a tuple in $\VIEW{R}$ from $\RINGZERO$ to non-$\RINGZERO$ (or vice versa), we increase (decrease) the count corresponding to the given $\mathcal{A}$-value. If this count changes from $0$ to $1$ (meaning the $\mathcal{A}$-value is unique) or from $1$ to $0$ (meaning there are no more tuples with the $\mathcal{A}$-value), then $\delta{(\VEXISTS{\mathcal{A}}\VIEW{R})}$ contains a tuple of $\mathcal{A}$-values with the payload of $\RINGONE$ or $-\RINGONE$, respectively; otherwise, the delta is empty.

\begin{example}\em
Consider a relation $\VIEW{R}$ over the schema $\{A,B\}$ with payloads from a ring $(\RING, \RINGPLUS, \RINGPROD, \RINGZERO, \RINGONE)$. We want to maintain the result of the query $\VIEW[A]{Q} = \VEXISTS{A}\VIEW[A,B]{R}$. To compute $\VIEW[A]{\delta{Q}}$ for updates to $\VIEW{R}$ efficiently, we count the tuples from $\VIEW{R}$ with non-$\RINGZERO$ payloads for each $A$-value, denoted by $\VIEW[A]{CNT_{Q}}$. For example:
\begin{small}
\begin{align*}
  \begin{tabular}[t]{@{}l|@{~}c@{~}c@{~}c@{~}c@{}}
    $\VIEW{R}$ & A & B \\
    \midrule
    & $a_1$ & $b_1$ & $\to$ & $r_1$ \\
    & $a_1$ & $b_2$ & $\to$ & $r_2$ \\  
    & $a_2$ & $b_3$ & $\to$ & $r_3$ \\
  \end{tabular}
  \quad
  \begin{tabular}[t]{@{}l|@{~}c@{~}c@{~}c@{}}
    $\VIEW{CNT_{Q}}$ & A \\
    \midrule
    & $a_1$ & $\to$ & $2$ \\
    & $a_2$ & $\to$ & $1$ \\
  \end{tabular}
  \quad
  \begin{tabular}[t]{@{}l|@{~}c@{~}c@{~}c@{}}
    $\VIEW{Q}$ & A \\
    \midrule
    & $a_1$ & $\to$ & $\RINGONE$ \\
    & $a_2$ & $\to$ & $\RINGONE$ \\
  \end{tabular}
\end{align*}
\end{small}
where $r_1$, $r_2$, and $r_3$ are non-$\RINGZERO$ payloads from $\RING$.
An update $\VIEW{\delta{R}} = \{ \tuple{a_1, b_2} \to -r_2 \}$ removes the tuple $\tuple{a_1,b_2}$ from $\VIEW{R}$, which in turn decreases $\VIEW[\text{$a_1$}]{CNT_{Q}}$ by $1$. Since there is still a tuple in $\VIEW{R}$ that projects on $a_1$, the result of $\VIEW{Q}$ remains unchanged. 
A subsequent update $\{ \tuple{a_1, b_1} \to -r_1 \}$ to $\VIEW{R}$ drops the count for $a_1$ to $0$, which triggers a change in the output, $\VIEW{\delta{Q}} = \{ \tuple{a_1} \to -\RINGONE \}$.
\punto
\end{example}

\begin{figure}[t]
\centering
\setlength{\tabcolsep}{3pt}
\begin{tabular}{@{}c@{}c@{~~}l@{}}
    \toprule
    \multicolumn{3}{c}{$I$ (\text{view tree} $\tau$)}\\
    \midrule
    \multicolumn{3}{l}{\MATCH $\tau$:}\\
    \midrule
    \phantom{ab} &
    \begin{minipage}[t]{3cm}
        \vspace{-1.6cm}
        \begin{tikzpicture}[xscale=0.45, yscale=1]
            \node at (0,-2)  (n4) {$\VIEW[{\it keys}]{V^{@X}_{rels}}$};
            \node at (-1,-3)  (n1) {$\tau_1$} edge[-] (n4);
            \node at (0,-3)  (n2) {$\ldots$};
            \node at (1,-3)  (n3) {$\tau_k$} edge[-] (n4);
        \end{tikzpicture}        
    \end{minipage}
    &
    \begin{minipage}[t]{9.5cm}
        \begin{tikzpicture}[xscale=1.0, yscale=1]
            \node at (0,-2)  (n4) {$\VIEW[{\it keys}]{V^{@X}_{rels}}$};
            \node at (-2,-3)  (n1) {$I(\tau_1)$} edge[-] (n4);
            \node at (-1.25,-3)  (n2) {$\ldots$};
            \node at (-0.5,-3)  (n1) {$I(\tau_k)$} edge[-] (n4);
            \node[color=red] at (0.5,-3)  (n1) {$\pi_1$} edge[red,dashed] (n4);
            \node[color=red] at (1.25,-3)  (n2) {$\ldots$};
            \node[color=red] at (2,-3)  (n3) {$\pi_l$} edge[red,dashed] (n4);
            \node at (3,-2.8) {$\text{  where}$};  
        \end{tikzpicture} \\[1ex]
        $\LET\SPACE \mathcal{R} \text{ be the set of all relation symbols }$\\[0.5ex]
        $\mathit{inds} = \{\, \VEXISTS{pk}\VIEW{R} \mid \VIEW{R} \in \mathcal{R} \setminus \VIEW{rels}, \mathit{pk} = \sch(\VIEW{R}) \cap \mathit{keys}, \mathit{pk} \neq \emptyset \,\}$\\[0.5ex]
        $\mathit{incycle} = \textsc{GYO}(\mathit{inds} \cup \{ \tau_1, \ldots, \tau_k \})$\\[0.5ex]
        $\{ \pi_1, \ldots, \pi_l \} = \mathit{incycle} \setminus \{ \tau_1, \ldots, \tau_k \}$
    \end{minipage}
    \\
    \bottomrule
\end{tabular}
\caption{Adding indicator projections to a view tree $\tau$. }
\label{fig:indicator_projections_algo}
\end{figure}
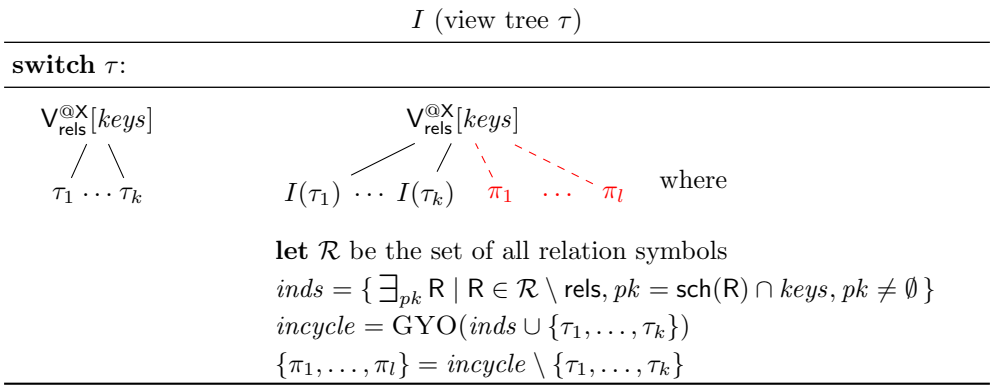

{\bf View Trees with Indicator Projections.}
Figure~\ref{fig:indicator_projections_algo} gives an algorithm that traverses a given view tree bottom-up and extends each view definition with indicator projections. At each view, the algorithm computes a set of relations $\mathit{inds}$ that could be used as indicator projections, restricting to only those relations that share common variables with that view and that do not appear in its definition. From this set of candidates, only those relations that form a cycle with the children of the given view are used as indicator projections. The algorithm uses the GYO reduction (Fagin et al variant)~\cite{GYO:1982} to determine this set of relations, denoted by $\mathit{incycle}$, and extends the view definition with the indicator projections of the candidates from this set. 

In a view tree with indicator projections, changes in one relation may propagate along multiple leaf-to-root paths. We propagate them in sequence, that is, updates to one relation are followed by a sequence of updates to its indicator projections. 

\begin{example}\em
The algorithm from Figure~\ref{fig:indicator_projections_algo} extends the view tree of the triangle query with an indicator projection $\VEXISTS{A,B}\VIEW[A,B]{R}$ placed below the view $\VIEW{V^{@C}_{ST}}$. This view at $C$ is now a cyclic join of the three relations, which can be computed in $\bigO{N^{3/2}}$ time. The indicator projection also reduces the size of this view to $\bigO{N}$.

Single-tuple updates to $S$ and $T$ still take linear time; however, bulk updates of size $\bigO{N}$ can now be processed in $\bigO{N^{3/2}}$ time, same as reevaluation. Updates to $R$ might affect the indicator projection: If a single-tuple update $\VIEW{\delta{R}}$ causes no change in the projection, then incremental maintenance takes constant time; otherwise, joining a tuple $\delta({\VEXISTS{A,B}\VIEW{R}})$ with $\VIEW{S}$ and $\VIEW{T}$ at node $C$ takes linear time. Bulk updates $\VIEW{\delta{R}}$ of size $\bigO{N}$ can also be processed in $\bigO{N^{3/2}}$ time. We conclude that using indicator projections in this query takes the best of both approaches from Example~\ref{ex:triangle_query_ivm}, namely faster incremental maintenance and more succinct view representation.
\punto
\end{example}


\section{Applications}
\label{sec:applications}

Our IVM framework supports a wide range of application scenarios. In this section, we highlight two scenarios, matrix chain computation and gradient computation for learning linear regression models over joins,
in which the payloads have sizes independent of the input relation sizes. We also highlight two distinct scenarios, where the payloads may have arbitrarily large sizes: they can be entire relations under either listing or factorized representations. 
{\em All these scenarios are treated uniformly using delta view trees
over the space of the view keys. They differ however in the rings used to define the view payloads.}

\subsection{Matrix Chain Multiplication}
\label{sec:mcm}

Consider the problem of computing a product of a series of matrices $\bm{A}_1, \ldots, \bm{A}_n$ over some ring $\RING$, where matrix $\bm{A}_i[x_i, x_{i+1}]$ has the size $p_{i} \times p_{i+1}$, $i \in [n]$. The product $\bm{A} = \bm{A}_1 \cdots \bm{A}_n$ is a matrix of size $p_1 \times p_{n+1}$ and can be formulated as follows:

\begin{align*}
\bm{A}[x_1, x_{n+1}] = \sum_{x_2 \in[p_2]} \cdots \sum_{x_n\in [p_n]} \prod_{i\in[n]} \bm{A}_i[x_i, x_{i+1}]
\end{align*}

We model a matrix $\bm{A}_i$ as a relation $\VIEW[X_i, X_{i+1}]{A_i}$ with the payload carrying matrix values. The query that computes the matrix $\bm{A}$ is:
\begin{align*}
\VIEW[X_1, X_{n+1}]{A} = \VSUM_{X_2} \cdots \VSUM_{X_n} \VPRODBIG_{i \in [n]} \VIEW[X_i, X_{i+1}]{A_i}
\end{align*}
where each of the lifting functions $\{g_{X_j}\}_{j \in [2,n]}$ maps any key value to payload $\RINGONE\in\RING$.
Different variable orders lead to different evaluation plans for matrix chain multiplication. The optimal variable order corresponds to the optimal sequence of matrix multiplications that minimizes the overall multiplication cost, which is the textbook Matrix Chain Multiplication problem~\cite{Cormen:2009:Algorithms}.

\begin{example}\em
\label{ex:MCM-factorized-update}
Consider a multiplication chain of $4$ matrices of equal size $p \times p$ represented as relations $\VIEW[X_i, X_{i+1}]{A_i}$. Let $\mathcal{F} = \{ X_1, X_5 \}$ be the set of free variables and $\omega$ be the variable order $X_1 - X_5 - X_3 - \{ X_2, X_4 \}$, i.e., $X_2$ and $X_4$ are children of $X_3$, with the matrix relations placed below the leaf variables in $\omega$. The view tree $\tau(\omega, \mathcal{F})$ has the following views (from bottom to top; the views at $X_5$ and $X_1$ are equivalent to the view at $X_3$):
\begin{align*}
\VIEW[X_1,X_3]{V^{@X_2}_{A_1A_2}} &= \textstyle\VSUM_{X_2} \VIEW[X_1,X_2]{A_1} \VPROD \VIEW[X_2,X_3]{A_2} \\
\VIEW[X_3,X_5]{V^{@X_4}_{A_3A_4}} &= \textstyle\VSUM_{X_4} \VIEW[X_3,X_4]{A_3} \VPROD \VIEW[X_4,X_5]{A_4} \\
\VIEW[X_1,X_5]{V^{@X_3}_{A_1A_2A_3A_4}} &= \textstyle\VSUM_{X_3} \VIEW[X_1,X_3]{V^{@X_2}_{A_1A_2}} \VPROD \VIEW[X_3,X_5]{V^{@X_4}_{A_3A_4}}
\end{align*}
Recomputing these views from scratch for each update to an input matrix takes $\bigO{p^3}$ time. A single-value change in any input matrix causes changes in one row or column of the parent view, and propagating them to compute the final delta view takes $\bigO{p^2}$ time. 
Updates to $\VIEW{A_2}$ and $\VIEW{A_3}$ change every value in $\VIEW{A}$. 
In case of a longer matrix chain, further propagating $\VIEW{\delta{A}}$ would require $\bigO{p^3}$ matrix multiplications, same as recomputation.

We exploit factorization to contain the computational effect of such changes. For instance, if $\VIEW{\delta{A_2}}$ is a factorizable update (see Section~\ref{sec:factorizable_updates}) expressible as $\VIEW[X_2,X_3]{\delta{A_2}} = \VIEW[X_2]{u} \VPROD \VIEW[X_3]{v}$, then we can propagate deltas more efficiently, as products of subexpressions:
\begin{align*}
\VIEW[X_1,X_3]{\delta{V}^{@X_2}_{A_1A_2}} &= \underbrace{\left( \textstyle\VSUM_{X_2} \VIEW[X_1,X_2]{A_1} \VPROD \VIEW[X_2]{u} \right)}_{\VIEW[X_1]{u_2}} \VPROD \VIEW[X_3]{v} \\
\VIEW[X_1,X_5]{\delta{V}^{@X_3}_{A_1A_2A_3A_4}} &= \VIEW[X_1]{u_2} \VPROD \left( \textstyle\VSUM_{X_3} \VIEW[X_3]{v} \VPROD \VIEW[X_3,X_5]{V^{@X_4}_{A_3A_4}} \right)
\end{align*}
Using such factorizable updates enables IVM in $\bigO{p^2}$ time.  The final delta is also in factorized form, suitable for further propagation. 

In general, for a chain of $k$ matrices of size $p \times p$, using a binary view tree of the lowest depth, incremental maintenance with factorizable updates takes $\bigO{p^2\log{k}}$ time, while reevaluation takes $\bigO{p^3 k}$ time. The space needed in both cases is $\bigO{p^2 k}$.
\punto
\end{example}

The above example recovers the main idea of LINVIEW~\cite{NEK:SIGMOD:2014}: exploit factorization for incremental computation of linear algebra programs when matrix changes are expressed as vector outer products, $\delta{A} = u \TR{v}$. Such rank-$1$ updates can capture many practical update patterns such as perturbations of one complete row or column in a matrix, or even changes of the whole matrix when the same vector is added to every row or column. Our framework generalizes this idea to arbitrary join-aggregate queries. 

\subsection{Gradient Computation}
\label{sec:application-lr}

In this section we introduce a ring that captures the computation of the gradient for training linear regression models over joins.

Consider a training dataset that consists of $k$ examples with $(X_i)_{i\in[m-1]}$ features/variables and a label $X_m$ arranged into a design matrix ${\bf M}$ of size $k \times m$; in our setting, this design matrix is the result of a join query. The goal of linear regression is to learn the model parameters $\Th = \TR{[\theta_1 \ldots \theta_m]}$ of a linear function\footnote{We consider wlog: $\theta_1$ is the bias parameter and then $X_1=1$ for all tuples in the input data; $\theta_m$ remains fixed to $-1$ and corresponds to the label/response $X_m$ in the data.} $f(X_1,\ldots,X_{m-1}) = \sum_{i\in[m]}\theta_iX_i$ best satisfying ${\bf M} \Th \approx {\bf 0}_{k\times 1}$, where ${\bf 0}_{k\times 1}$ is the $k$-by-$1$ matrix with all elements 0.

We can solve this optimization problem using batch gradient descent. This method iteratively updates the model parameters in the direction of the gradient to decrease the squared error loss and eventually converge to the optimal value. Each convergence step iterates over the entire training dataset to update the parameters, $\Th := \Th - \alpha\TR{\bf M}{\bf M}\Th$, where $\alpha$ is an adjustable step size. The complexity of each step is $\bigO{mk}$. The {\em cofactor matrix}  $\TR{\bf M}{\bf M}$ quantifies the degree of correlation for each pair of features (or feature and label). Its computation is dependent on the data and can be executed once for all convergence steps~\cite{SOC:SIGMOD:2016}. This is crucial for performance in case $m \ll k$ as each iteration step now avoids processing the entire training dataset and takes time $\bigO{m^2}$. 

\nop{
Computing the cofactor matrix $\TR{\X}\X$, which quantifies the degree of correlation for each pair of features, can be expensive. When $\X$ is the result of an equi-join query $Q$ over database $\db$, computing $\TR{\X}\X$ taking a flat join result as input would take $\bigO{m^2 \cdot |\db|^{\rho^{*}(Q)}}$ time. But computing the cofactor matrix over a factorized join result takes $\bigO{m^2 \cdot |\db|^{\mathit{fhtw}(Q)}}$ time, where the gap between $\rho^{*}(Q)$ and $fhtw(Q)$ can be as big as the number of relations in $Q$.
}

We next show how to incrementally maintain the cofactor matrix. We can factorize its computation over training datasets defined by results to arbitrary join queries~\cite{SOC:SIGMOD:2016}. The idea is to compute a triple of regression aggregates $(\LRringC,\LRringS,\LRringQ)$, where $\LRringC$ is the number of tuples in the training dataset (size $k$ of the design matrix), $\LRringS$ is an $m\times 1$ matrix (or vector) with one sum of values per variable, and $\LRringQ$ is an $m\times m$ matrix of sums of products of values for any two variables. Their incremental computation can be captured by a ring.
\begin{definition}\em
\label{def:lr_ring}
For a fixed $m \in \mathbb{N}$, let $\RING$ denote the set of triples $(\mathbb{Z}, \mathbb{R}^{m}, \mathbb{R}^{m \times m})$, $\RINGZERO = (0, {\bf 0}_{m \times 1}, {\bf 0}_{m \times m})$ and $\RINGONE = (1, {\bf 0}_{m \times 1}, {\bf 0}_{m \times m})$. For $a = (\LRringC_a, \LRringS_a, \LRringQ_a) \in \RING$ and $b = (\LRringC_b, \LRringS_b, \LRringQ_b) \in \RING$, define the operations $+^{\RING}$ and $*^{\RING}$ over $\RING$ as:
\begin{align*}
a +^{\RING} b &= (\LRringC_a + \LRringC_b, \LRringS_a + \LRringS_b, \LRringQ_a + \LRringQ_b) \\
a *^{\RING} b &= (\LRringC_a \LRringC_b, \LRringC_b \LRringS_a + \LRringC_a \LRringS_b, \LRringC_b \LRringQ_a + \LRringC_a \LRringQ_b + \LRringS_a \TR{\LRringS_b} + \LRringS_b \TR{\LRringS_a}) 
\end{align*}
The structure $(\RING, +^{\RING}, *^{\RING}, \RINGZERO, \RINGONE)$ forms the {\em degree-$m$ matrix} ring.
\end{definition}

We next show how to use this ring to compute the cofactor matrix over a training dataset defined by a join query with relations $(\VIEW{R_i})_{i\in[n]}$ over variables $(X_j)_{j\in[m]}$. The payload of each tuple in a relation is the identity ${\bf 1}$ from the degree-$m$ matrix ring. The query expressing the computation of the cofactor matrix is:
\begin{align*}
\VIEW{Q} = \textstyle\VSUM_{X_1}{} \cdots \VSUM_{X_m}{\VPRODBIG_{i \in [n]} \VIEW[\mathit{\sch(R_i)}]{R_i}}
\end{align*}
For each $X_j$-value $x$, the lifting function is $g_{X_j}(x) = (1, \LRringS, \LRringQ)$, where $\LRringS$ is an $m \times 1$ vector with all zeros except the value of $x$ at position $j$, i.e., $\LRringS_j=x$, and $\LRringQ$ is an $m \times m$ matrix with all zeros except the value $x^2$ at position $(j,j)$: $\LRringQ_{(j,j)}=x^2$.

\nop{
To understand the intuition behind this ring, consider the cofactor matrix computation over a join result expressed as a matrix $\X$. 
Then, $\LRringC$ corresponds to the total number of tuples in $\X$, $\LRringS$ contains the sum over each variable, and $\LRringQ$ is the cofactor matrix $\TR{\X}\X$.
Consider now two matrices $\X_1$ and $\X_2$ for disjoint partitions of the join result, and their regression aggregates $(\LRringC_1, \LRringS_1, \LRringQ_1)$ and $(\LRringC_2, \LRringS_2, \LRringQ_2)$. Then, $\X = \TR{\begin{bmatrix} \X_1 \; \X_2 \end{bmatrix}}$, the aggregates for $\X$ are:
\begin{align*}
\LRringC \;=&\; \LRringC_1 + \LRringC_ 2 \\
\LRringS \;=&\; \mathit{sum}(\X) = \mathit{sum}(\X_1) + \mathit{sum}(\X_2) = \LRringS_1 + \LRringS_2 \\
\LRringQ \;=&\; \TR{\X} \X = \TR{\X_1} \X_1  + \TR{\X_2} \X_2 = \LRringQ_1 + \LRringQ_2
\end{align*}
where $\mathit{sum}$ returns a vector with the sum of each column.
Now consider $\X_1$ and $\X_2$ for two vertical partitions whose product is the join result. Then, $\X$ has $\X_1$ duplicated $\LRringC_2$ times and $\X_2$ duplicated $\LRringC_1$ times, hence the rescaling of the sum and quadratic aggregates in the definition of $*^{\RING}$. The product of $\X_1$ and $\X_2$ also forms new interactions between features from different datasets, captured via the product of their linear aggregates.
}

\begin{example}\em
\label{ex:gradient-computation}
We show how to compute the cofactor matrix over the join, database, and view tree from Figure~\ref{fig:example_viewtree}. We assume alphabetical order over the five variables. The leaves $\VIEW{R}$, $\VIEW{S}$, and $\VIEW{T}$ are the input relations that map tuples to $\RINGONE$ from the degree-$5$ matrix ring. 

In the view $\VIEW{V^{@D}_{T}}$ each $D$-value $d$ is lifted to a triple $(1, \LRringS, \LRringQ)$, where $\LRringS$ is a $5\times 1$ vector with one non-zero element $\LRringS_4=d$, and $\LRringQ$ is a $(5 \times 5)$ matrix with one non-zero element $\LRringQ_{(4,4)} = d^2$. Then, those regression triples with the same key $c$ are summed up, yielding $\VIEW{V^{@D}_{T}}[c_1]=(1,\LRringS_4=d_1, \LRringQ_{(4,4)}=d_1^2)$, $\VIEW{V^{@D}_{T}}[c_2]=(2,\LRringS_4=d_2+d_3, \LRringQ_{(4,4)}=d_2^2+d_3^2)$,
and $\VIEW{V^{@D}_{T}}[c_3]=(1,\LRringS_4=d_4, \LRringQ_{(4,4)}=d_4^2)$. The views $\VIEW{V^{@B}_{R}}$ and $\VIEW{V^{@E}_{S}}$ are computed in a similar way.

The view $\VIEW{V^{@C}_{ST}}$ joins the views $\VIEW{V^{@D}_{T}}$ and $\VIEW{V^{@E}_{S}}$ and then marginalizes $C$. For instance, the payload for the key $\VIEW{V^{@C}_{ST}}[a_2]$ is as follows:
{\setlength{\arraycolsep}{1pt}
\begin{align*}
&\VIEW{V^{@C}_{ST}}[a_2] = \VIEW[\mathit{c_2}]{V^{@D}_{T}} *^{\RING} \VIEW[\mathit{a_2, c_2}]{V^{@E}_{S}} *^{\RING} g_{C}(c_2) \\[0.5ex]
&=
\left(\!
    2,\!
    \begin{bmatrix}
    0 \\ 0 \\ 0 \\ d_2 \!+\! d_3 \\ 0
    \end{bmatrix}\!\!,\!
    \begin{bmatrix}
    0 & 0 & 0 & 0 & 0 \\
    0 & 0 & 0 & 0 & 0 \\
    0 & 0 & 0 & 0 & 0 \\
    0 & 0 & 0 & d_2^2 \!+\! d_3^2 & 0 \\
    0 & 0 & 0 & 0 & 0
    \end{bmatrix}
\!\right) 
\!*^{\RING}\!
\left(\!
    1,\!
    \begin{bmatrix}
    0 \\ 0 \\ 0 \\ 0 \\ e_4
    \end{bmatrix}\!\!,\!
    \begin{bmatrix}
    0 & 0 & 0 & 0 & 0 \\
    0 & 0 & 0 & 0 & 0 \\
    0 & 0 & 0 & 0 & 0 \\
    0 & 0 0 & 0 & 0 \\
    0 & 0 0 & 0 & e_4^2
    \end{bmatrix}
\!\right) 
\!*^{\RING}\!
\left(\!
    1,\!
   \begin{bmatrix}
     0 \\ 0 \\ c_2 \\ 0 \\ 0
    \end{bmatrix}\!\!,\!
    \begin{bmatrix}
    0 & 0 & 0 & 0 & 0 \\
    0 & 0 & 0 & 0 & 0 \\
    c_2^2 & 0 & 0 & 0 & 0\\
    0 & 0 & 0 & 0 & 0 \\
    0 & 0 & 0 & 0 & 0
    \end{bmatrix}
\!\right)\\[0.5ex]
&= 
\left(
  2,
  \begin{bmatrix}
    0 \\ 0 \\ 2 c_2 \\ d_2 + d_3 \\ 2 e_4
  \end{bmatrix},
  \begin{bmatrix}
    0 & 0 & 0 & 0 & 0 \\
    0 & 0 & 0 & 0 & 0 \\
    0 & 0 & 2c_2^2 & c_2(d_2 + d_3) & 2 c_2 e_4 \\
    0 & 0 & c_2(d_2 + d_3) & d_2^2 + d_3^2 & (d_2 + d_3)e_4 \\
    0 & 0 & 2 c_2 e_4 & (d_2 + d_3)e_4 & 2 e_4^2
  \end{bmatrix}
\right)
\end{align*}
}

The root view $\VIEW{V^{@A}_{RST}}$ maps the empty tuple to the ring element $\sum_{l\in[2]}\VIEW[a_l]{V^{@B}_{R}} *^{\RING} \VIEW[a_l]{V^{@C}_{ST}} *^{\RING} g_{A}(a_l)$.
This payload has aggregates for the entire join result: the count of tuples in the result, the vector with one sum of values per variable, and the cofactor matrix.
\punto
\end{example}

For performance reasons, in practice we only store as payloads blocks of matrices with non-zero values and assemble larger matrices as the computation progresses towards the root of the view tree. We can further exploit the symmetry of the cofactor matrix to compute only the entries above and including the diagonal.



\subsection{Factorized Representation of Query Results}
\label{sec:relational-ring}

Our framework can also support scenarios where the view payloads are themselves relations representing results of conjunctive queries, or even their factorized representations. Factorized representations can be arbitrarily smaller than the listing representation of a query result~\cite{Olteanu:FactBounds:2015:TODS}, with orders of magnitude size gaps reported in practice~\cite{SOC:SIGMOD:2016}. They nevertheless remain lossless and support constant-delay enumeration of the tuples in the query result as well as subsequent aggregate processing in one pass. Besides the factorized view computation and the factorizable updates, this is the third instance where our framework exploits factorization.

We first introduce the relational data ring that allows us to store entire relations as payloads. We then show how to encode factorized representation of relations in payloads. 
By using a relational data ring, we can create payloads that hold relations. When marginalizing a variable, we move its values from the key space to the payload space. The payloads of the tuples of a view are now relations over the same schema. These relations have themselves payloads in the $\mathbb{Z}$ ring used to maintain the multiplicities of their tuples. \nop{This approach can be used to maintain conjunctive queries, as an alternative to the approach where the views keep the tuples in the keys and map them to multiplicities in the $\mathbb{Z}$ ring.}

\begin{definition}\em
Let $\mathbb{F}[\mathbb{Z}]$ denote the set of relations over the $\mathbb{Z}$ ring, 
the zero $\RINGZERO$ in $\mathbb{F}[\mathbb{Z}]$ is the empty relation $\{\}$, which maps every tuple to  $0\in\mathbb{Z}$, and the identity ${\bf 1}$ is the relation $\{ () \rightarrow 1 \}$, which maps the empty tuple to $1 \in\mathbb{Z}$ and all other tuples to $0 \in\mathbb{Z}$. The structure $(\mathbb{F}[\mathbb{Z}], \VPLUS, \VPROD, \RINGZERO, \RINGONE)$ forms the {\em relational data} ring.\footnote{
To form a proper ring, we would need a generalization~\cite{Koch:Ring:2010:PODS} of our relations and join and union operators, where: tuples have their own schemas; union may apply to tuples with possibly different schemas; join accounts for multiple derivations of output tuples. For our practical needs this generalization is not necessary.}
\end{definition}

We model conjunctive queries as count queries that marginalize {\em every} variable but use different lifting functions for the free and bound variables. For a variable $X$ and any of its values $x$, $g_{X}(x) = \{ (x) \to 1 \}$ if $X$ is a free variable and $g_{X}(x) = \RINGONE = \{ () \to 1 \}$ if $X$ is bound; here, $(x)$ is a singleton relation  over schema $\{X\}$. We have relational operations occurring at two levels: for keys, we join views and marginalize variables as before; for payloads, we interpret multiplication and addition of payloads as join and union of relations.
\begin{example}\em
\label{ex:relational_ring}
Consider the conjunctive query
\begin{align*}
Q(A,B,C,D) = R(A,B), S(A,C,E), T(C,D)
\end{align*}
over the three relations from Figure~\ref{fig:example_payloads}, where each tuple gets the identity payload $\{ \tuple{} \to 1 \} \in \mathbb{F}[\mathbb{Z}]$. The corresponding view is:
\begin{align*}
 \VIEW[~]Q = \textstyle\VSUM_{A}\VSUM_{B}\VSUM_{C}\VSUM_{D}\VSUM_{E} \VIEW[A,B]{R} \VPROD \VIEW[A,C,E]{S} \VPROD \VIEW[C,D]{T}
\end{align*}
The lifting function for $E$ maps each value to $\{() \to 1 \}$, while the lifting functions for all other variables map value $x$ to $\{(x) \to 1 \}$.

Figure~\ref{fig:example_payloads} shows a view tree for this query and the contents of its views with relational data payloads (in black and red) for the given database. The view keys gradually move to payloads as the computation progresses towards the root. The view definitions are identical to those of the {\tt COUNT} query (but under a different ring!). The view $\VIEW{V^{@D}_{T}}$ lifts each $D$-value $d$ from $\VIEW{T}$ to the relation $\{ (d) \to 1 \}$ over schema $\{D\}$, multiplies (joins) it with the payload $\RINGONE$ of each tuple, and sums up (union) all payloads with the same $c$-value. The views at $\VIEW{V_R^{@B}}$ and $\VIEW{V_S^{@E}}$ are computed similarly, except the latter lifts $e$-values to $\{ \tuple{} \to 1 \}$ since $E$ is a bound variable. 
The view {\color{red}$\VIEW{V^{@C}_{ST}}$} assigns to each $A$-value a payload that is a union of Cartesian products of the payloads of its children and the lifted $C$-value. The root view {\color{red}$\VIEW{V^{@A}_{RST}}$} similarly computes the payload of the empty tuple, which represents the query result (both views are at the right).
\punto
\end{example}

We next show how to construct a factorized representation of the query result. In contrast to the scenarios discussed above, this representation is {\em not} available as one payload at the root view, but {\em distributed} over the payloads of all views. This hierarchy of payloads, linked via the keys of the views, becomes the factorized representation. A further difference lies with the multiplication operation. For the listing representation, the multiplication is the Cartesian product. For a given view, it is used to concatenate payloads from its child views. For the factorized representation, we further project away values for all but the marginalized variable. More precisely, for each view $\VIEW[\mathcal{S}]{V^{@X}_{rels}}$ and each of its keys $a_{\mathcal{S}}$, let $\VIEW[\mathcal{T}]{P} = \VIEW[a_\mathcal{S}]{V^{@X}_{rels}}$ be the corresponding payload relation. Then, instead of computing this payload, we compute $\VSUM_{Y\in \mathcal{T}-\{X\}}\VIEW[\mathcal{T}]{P}$ by marginalizing the variables in $\mathcal{T}-\{X\}$ and summing up the multiplicities of the tuples in $\VIEW[\mathcal{T}]{P}$ with the same $X$-value.

\begin{example}\em
\label{ex:factorized_ring}
We continue Example~\ref{ex:relational_ring}. Figure~\ref{fig:example_payloads_conjunctive} shows the contents of the views with factorized payloads (in black and blue). Each view stores relational payloads that have the schema of the marginalized variable. 
Together, these payloads form a factorized representation over the variable order $\omega$ used to define the view tree in Figure~\ref{fig:example_payloads}. At the top of the factorization, we have a union of two $A$-values: $a_1$ and $a_2$. This is stored in the payloads of (middle) {\color{blue}$\VIEW[\;]{V^{A}_{RST}}$}. The payloads of (middle) {\color{blue}$\VIEW[A]{V^{@C}_{ST}}$} store a union of $C$-values $c_1$ and $c_2$ under $a_1$, and a singleton union of $c_2$ under $a_2$. The payloads of $\VIEW[A]{V^{@B}_R}$ store a union of $B$-values $b_1$ and $b_2$ under $a_1$ and a singleton union of $b_3$ under $a_2$. Note the (conditional) independence of the variables $B$ and $C$ given a value for $A$. This is key to succinctness of factorization. In contrast, the listing representation explicitly materializes all pairings of $B$ and $C$-values for each $A$-value, as shown in the payload of (right) {\color{red}$\VIEW[\;]{V^{A}_{RST}}$}. Furthermore, the variable $D$ is independent of the other variables {\em given} $C$. This is a further source of succinctness in the factorization: Even though $c_2$ occurs under both $a_1$ and $a_2$, the relations under $c_2$, in this case the union of $d_2$ and $d_3$, is only stored once in $\VIEW[C]{V^{@D}_T}$. Each value in the factorization keeps a multiplicity, that is, the number of its derivations from the input data. This is necessary for maintenance. 

This factorization is over a variable order that can be used for all queries with same body and different free variables: As long as their free variables sit on top of the bound variables, the variable order is valid and so is the factorization over it. For instance, if the variable $D$ were not free,  then the factorization for the new query would be the same except that we now discard the unions of $D$-values.\punto
\end{example}


\section{Experiments}
\label{sec:experiments}

We compare \DF (factorized IVM) against \IVM (first-order IVM) and \DBT (DBToaster's fully recursive higher-order IVM). Our experimental results can be summarized as follows:
\begin{itemize}
    \item Factorized updates lead to two orders of magnitude speedup for \DF over competitors for matrix chain multiplication by propagating factorized deltas and avoiding matrix multiplication.
    \item For cofactor matrices used in regression models, \DF exhibits the lowest memory utilization and up to two orders of magnitude better performance than \IVM and \DBT.  

    \item 
    For conjunctive query evaluation, factorized payloads can speed up view maintenance and reduce memory by up to two orders of magnitude compared to the listing representation of payloads.
\end{itemize}

\subsection{Experimental Setting}

{\bf Runtime.}
\IVM and \DBT are supported by DBToaster~\cite{DBT:VLDBJ:2014}, a system that compiles a given SQL query into code that maintains the query result under updates to input relations. The generated code represents an in-memory stream processor that is standalone and independent of any database system. DBToaster's performance on decision support and financial workloads can be several orders of magnitude better than state-of-the-art commercial databases and stream processing systems~\cite{DBT:VLDBJ:2014}. We implemented \DF as a program that maintains a set of materialized views for a given variable order and a set of updatable relations. We use the intermediate language of DBToaster to encode this program and then feed it into DBToaster's code generator. We modified the backend of DBToaster v2.2 to enable arbitrary ring payloads and limit the amount of memory over-provisioning to at most one million records.
Unless stated otherwise, all the benchmarked approaches use the same runtime and store views as multi-indexed maps with memory-pooled records. The algorithms and record types used in these approaches, however, can differ greatly.

{\bf Workload. }
We run experiments over three datasets:

\begin{itemize}[leftmargin=0mm,itemindent=26pt]
    \item {\em Retailer} is a real-world dataset from an industrial collaborator and used by a retailer for business decision support and forecasting user demands. 
    The dataset has a snowflake schema with one fact relation {\tt Inventory} with $84$M records, storing information about the inventory units for products in a location, at a given date.
    The {\tt Inventory} relation joins along three dimension hierarchies: {\tt Item} (on product id), {\tt Weather} (on location and date), and {\tt Location} (on location) with its lookup relation {\tt Census} (on zip code). 
    The natural join of these five relations is acyclic and has $43$ attributes. We consider a view tree in which the variables of each relation form a distinct root-to-leaf path, and the partial order on join variables is: location - $\{$ date - $\{$ product id $\}$, zip $\}$.
     
    \item {\em Housing} is a synthetic dataset modeling a house price market~\cite{SOC:SIGMOD:2016}.
    It consists of six relations: {\tt House}, {\tt Shop}, {\tt Institution}, {\tt Restaurant}, {\tt Demographics}, and {\tt Transport}, arranged into a star schema and with $1.4$M tuples in total (scale factor 20). The natural join of all relations is on the common attribute (postcode) and has $27$ attributes. We consider an optimal view tree that has each root-to-leaf path consisting of query variables of one relation.

    \item {\em Twitter} represents friends/followers relationships among users who were active on Twitter during the discovery of Higgs boson~\cite{Higgs:TwitterDataset}. We split the first $3$M records from the dataset into three equally-sized relations, $R(A,B)$, $S(B,C)$, and $T(C,A)$, and consider the triangle query over them and the variable order $A - B - C$.
\end{itemize}

We run the systems over data streams synthesized from the above datasets by interleaving insertions to the input relations in a round-robin fashion. We group insertions into batches of different sizes and place no restriction on the order of records in input relations. 
In all experiments, we use payloads defined over rings with additive inverse, thus processing deletions is similar to that of insertions.

{\bf Queries.} 
We next present the queries used in the experiments. 

\begin{itemize}
\item 
{\em Matrix Chain Multiplication:}
The query in standard SQL is defined over tables $A_1(I,J,P_1)$, $A_2(J,K,P_2)$, $A_3(K,L,P_3)$:
\begin{lstlisting}[language=SQL,columns=flexible]
        SELECT A1.I, A3.L, SUM(A1.P1 * A2.P2 * A3.P3)
        FROM A1 NATURAL JOIN A2 NATURAL JOIN A3
        GROUP BY A1.I, A3.L;
\end{lstlisting}
In our formalism, each relation maps pairs of indices to matrix values, all lifting functions map values to $1$, and the query is: 
$\VIEW[I,L]{Q}=\VSUM_{J}\VSUM_{K} \VIEW[I,\textsf{$J$}]{A_1} \VPROD \VIEW[\textsf{$J$},K]{A_2} \VPROD \VIEW[K,L]{A_3}$.

\item 
{\em Cofactor Matrix Computation:}
For the {\em Retailer} schema, the query has one regression aggregate over the natural join of its relations:
\begin{lstlisting}[language=SQL,columns=flexible, mathescape]
        SELECT SUM(g$_1$(X$_1$) * ... * g$_{43}$(X$_{43}$))
        FROM Inventory NATURAL JOIN Item NATURAL JOIN Weather
                        $\;$NATURAL JOIN Location NATURAL JOIN Census;
\end{lstlisting}
where $\{ X_i \}_{i \in [43]}$ are all the variables from the {\em Retailer} schema, 
the {\tt SUM} operator uses $+$ and $*$ from the degree-$43$ matrix ring, 
and each lifting function $g_i$ maps a value $x$ to $g_i(x) = (\LRringC_i=1, \LRringS_i = x, \LRringQ_{(i,i)} = x^2$) (see Example~\ref{ex:gradient-computation}).
Similarly, the queries over {\em Housing} and {\em Twitter} use the degree-$27$ and respectively degree-$3$ matrix ring. 
We consider all variables to be continuous; categorical variables can be treated using group-by queries as explained in related work~\cite{ANNOS:PODS:2018}.

\item
{\em Factorized Computation of Conjunctive Queries:}
We consider two full conjunctive queries joining all the relations in the {\em Retailer} and respectively {\em Housing} datasets. 
\end{itemize}

{\bf Experimental Setup. }
We run all experiments on a Microsoft Azure DS14 instance, Intel(R) Xeon(R) CPU E5-2673 v3 @ 2.40GHz, 112GB RAM, with Ubuntu Server 14.04. 
We use DBToaster v2.2 for the IVM competitors and code generation in our approach. 
The generated C++ code is single-threaded and compiled using g++ 6.3.0 with the -O3 flag. 
We set a one-hour timeout on query execution and report wall-clock times by averaging three best results out of four runs. 
We profile memory utilization using gperftools, not counting the memory used for storing input streams.


\begin{figure}[t]
\begin{center}
{
\renewcommand{\arraystretch}{1.3}
\begin{tabular}{lccccc}
\toprule
 & \DF &  \DBT & \IVM & \DFRE & \DBTRE \\
\midrule
Retailer \quad & $2,955,045$ &  $1,250,262$ & $2,925,828$ & $3,785^{*}$ & $3,491^{*}$ \\
Housing  \quad & $22,857,143$ & $17,834,395$ & $2,403,433$ & $79,226$ & $364^{*}$ \\
\bottomrule
\end{tabular}
}
\end{center}
\caption{The average throughput (tuples/sec) of reevaluation and incremental maintenance of a sum aggregate under updates of size $1,000$ to all relations of the {\em Retailer} and {\em Housing} datasets with a one-hour timeout (denoted by the symbol$^{*}$).}
\label{table:sum_aggregate_cost_comparison}
\end{figure}

\subsection{Maintenance of Sum Aggregates}
We analyze different strategies for maintaining a sum of one variable on top of a natural join. We measure the average throughput of reevaluation and incremental maintenance under updates of size $1,000$ to all the relations of {\em Retailer} and {\em Housing}. For the former dataset, we sum the inventory units for products in {\tt Inventory}; for the latter, we sum over the common join variable. We also benchmark two reevaluation strategies that recompute the results from scratch on every update: \DFRE denotes reevaluation using variable orders and \DBTRE denotes reevaluation using DBToaster. Table~\ref{table:sum_aggregate_cost_comparison} summarizes the results.

\DF achieves the highest average throughput in both cases. For {\em Retailer}, the maintenance cost is dominated by the update on {\tt Inventory}. 
\DBT's recursive delta compilation materializes $13$ views representing connected subqueries: five group-by aggregates over the input relations, {\tt Inv}, {\tt It}, {\tt W}, {\tt L}, and {\tt C}; one group-by aggregate joining {\tt L} and {\tt C}; six views joining {\tt Inv} with subsets of the others, namely \{{\tt It}\}, \{{\tt It}, {\tt W}\}, \{{\tt It}, {\tt W}, {\tt L}\}, \{{\tt W}\}, \{{\tt W}, {\tt L}\}, and \{{\tt W}, {\tt L}, {\tt C}\}; and the final aggregate.
The two views joining {\tt Inv} with \{ {\tt W}, {\tt L} \} and \{ {\tt It}, {\tt W}, {\tt L} \} require linear maintenance for a single-tuple change in {\tt Inventory}.
\IVM recomputes deltas from scratch on each update using only the input relations with no aggregates on top of them. Updates to {\tt Inventory} are efficient due to small sizes of the other relations. 
\DF uses the given variable order to materialize $9$ views, four of them over {\tt Inventory}, \{{\tt Inv}\}, \{{\tt Inv}, {\tt It}\}, \{ {\tt Inv}, {\tt It}, {\tt W} \}, and the final sum, but each with constant maintenance for single-tuple updates to this relation.
In contrast to \IVM, our approach materializes precomputed views in which all nonjoin variables are aggregated away. 
In the {\em Housing} schema, both \DF and \DBT benefit from this preaggregation, and since the query is a star join, both materialize the same views. \DBT computes {\tt SUM(1)} and {\tt SUM(postcode)} for each {\tt postcode} in the delta for {\tt Inventory}, although only the count suffices.
Figure~\ref{table:sum_aggregate_cost_comparison} also shows that the reevaluation strategies significantly underperform the incremental approaches.


\begin{figure*}[t]
  \centering   
  \includegraphics[width=0.47\textwidth]{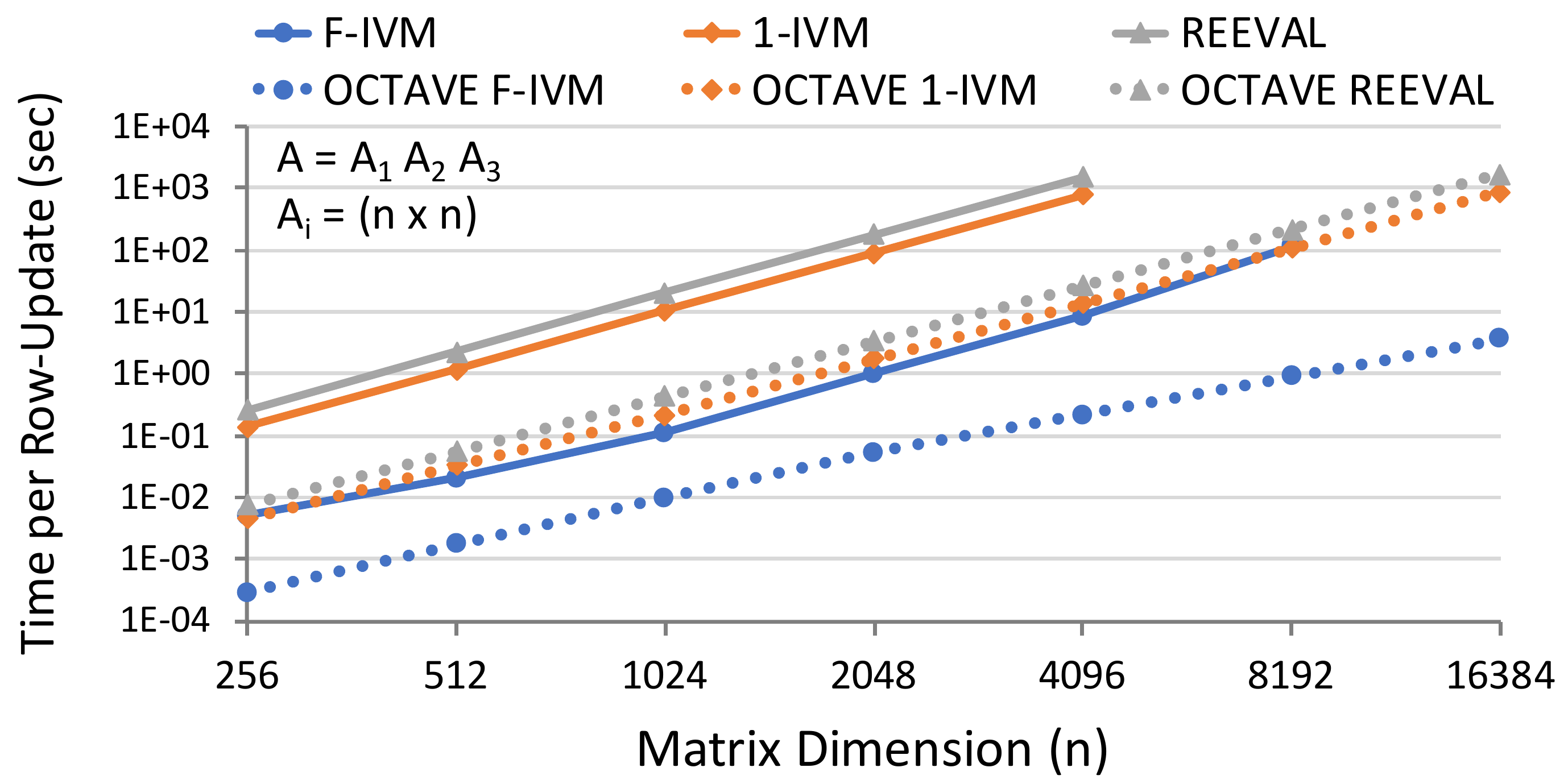}
  \qquad
  \includegraphics[width=0.47\textwidth]{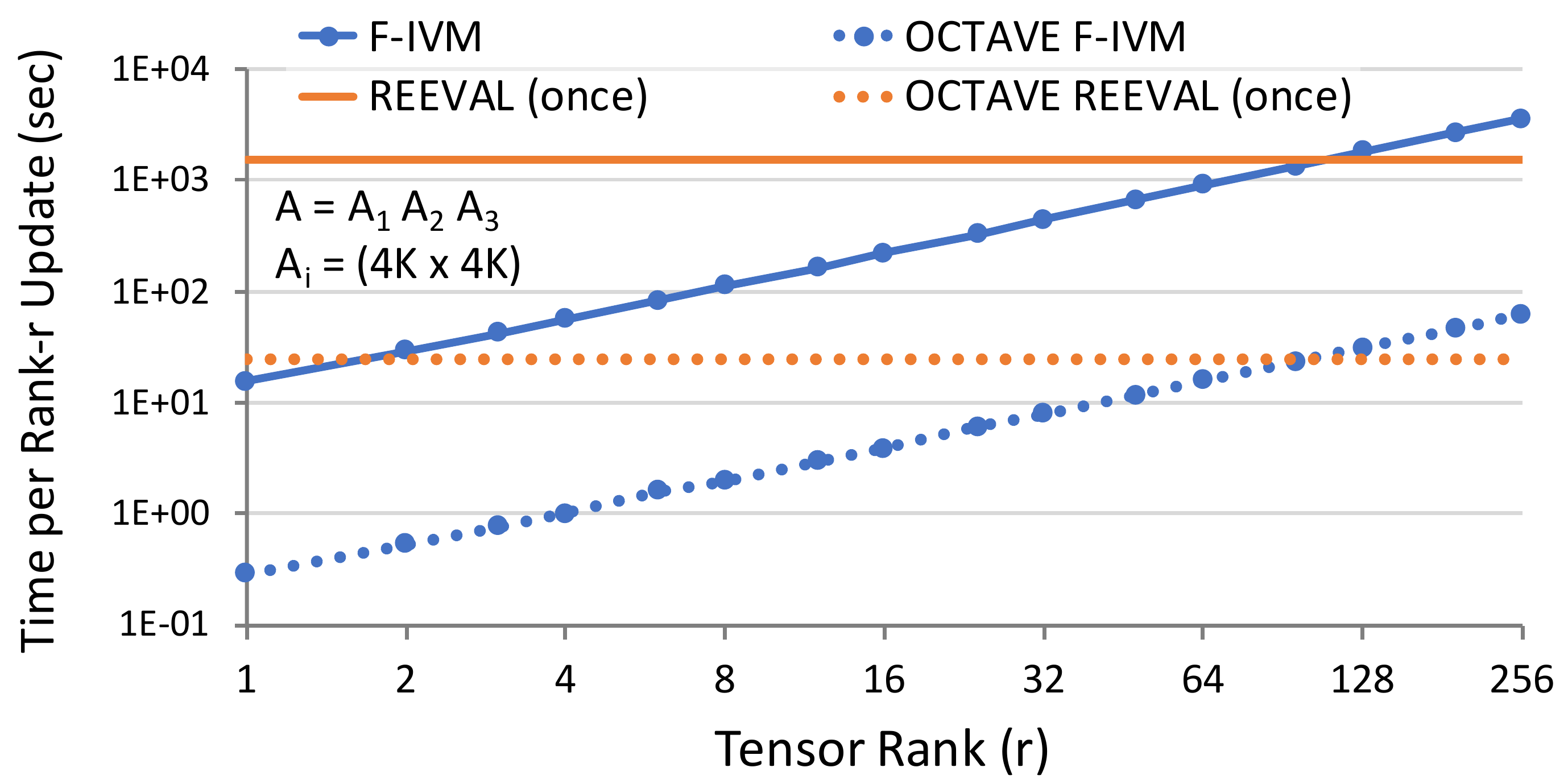}
  \caption{Incremental maintenance and reevaluation of the product of three $(n \times n)$ matrices, $A = A_1 \, A_2 \, A_3$: (left) one-row updates in $A_2$; (right) rank-$r$ updates in $A_2$ for $n=4,096$ using the DBToaster and Octave runtime environments. }
  \label{fig:MCM}
\end{figure*}

\subsection{Matrix Chain Multiplication with Factorized Updates}

We consider the problem of maintaining the multiplication $A = A_1 \, A_2 \, A_3$ of three $(n \times n)$ matrices under changes to $A_2$. We compare \DF with factorized updates, \IVM that recomputes the delta $\delta{A} = A_1 \, \delta{A_2} 
\, A_3$ from scratch, and REEVAL that recomputes the entire product from scratch on every update. 
\DBT becomes \IVM in this particular setting.
We consider two different implementations of these maintenance strategies: The first uses DBToaster's hash maps to store matrices, while the second uses Octave, a numerical tool that stores matrices in dense arrays and offers highly-optimized BLAS routines for matrix multiplication~\cite{Whaley1999}. In both cases, matrix-matrix multiplication takes $\bigO{n^{\alpha}}$ for $\alpha > 2$; for instance, $\alpha = 2.8074$ for 
Strassen's algorithm.

We first consider updates to one row in $A_2$. For \IVM, the delta $\delta{A_{12}} = A_1 \, \delta{A_2}$ might  contain non-zero changes to all $n^2$ matrix entries, thus computing $\delta{A} = \delta{A_{12}} \, A_3$ requires full matrix-matrix multiplication. REEVAL updates $A_2$ first before computing two matrix-matrix multiplications. \DF factorizes $\delta{A_2}$ into a product of two vectors $\delta{A_2} = u \TR{v}$, which are used to compute $\delta{A_{12}} = (A_1 \, u) \, \TR{v} = u_1 \, \TR{v}$ and $\delta{A} = u_1 \, (\TR{v} \, A_3) = u_1 \, v_1$. Both deltas involve only matrix-vector multiplications computed in $\bigO{n^2}$ time. Figure~\ref{fig:MCM} (left) shows the average time needed to process an update to one randomly selected row in $A_2$ for different matrix sizes. REEVAL performs two matrix-matrix multiplications, while \IVM performs only one. In the hash-based implementation, the gap between \DF and \IVM grows from $28$x for $n=256$ to $92$x for $n=4,096$; similarly, in the Octave implementation, the same gap grows from $16$x for $n=256$ to $236$x for $n=16,384$. This confirms the difference in the asymptotic complexity of these strategies.

Our next experiment considers rank-$r$ updates to $A_2$, which can be decomposed into a sum of $r$ rank-$1$ tensors, $\delta{A_2} = \sum_{i\in[r]} u_i \TR{v_i}$. \DF processes $\delta{A_2}$ as a sequence of $r$ rank-$1$ updates in $\bigO{rn^2}$ time, while both REEVAL and \IVM take as input one full matrix $\delta{A_2}$ and maintain the product in $\bigO{n^3}$ time per each rank-$r$ update. \IVM has the same performance as REEVAL. Figure~\ref{fig:MCM} (right) shows that the average time \DF takes to process a rank-$r$ update for different $r$ values and the matrix size $4,096$ is linear in the tensor rank $r$. 
Under both implementations in DBToaster and Octave, incremental computation is faster than reevaluation for updates with rank $r\leq 96$.
With larger matrix sizes, the gap between reevaluation and incremental computation increases, which enables incremental maintenance for updates of higher ranks.

\begin{figure*}[t]
  \centering   
  \includegraphics[width=0.47\textwidth]{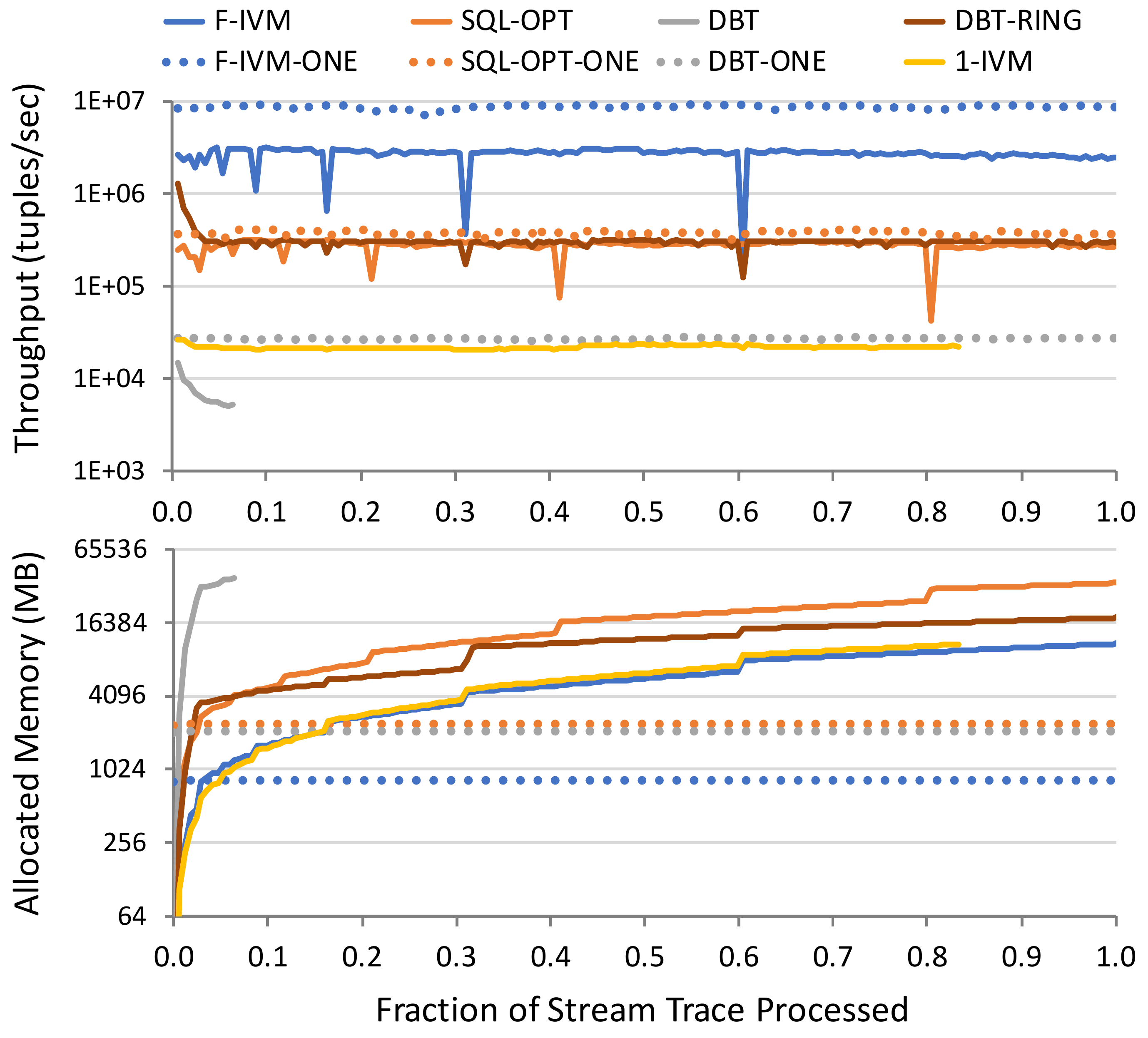}
  \quad\;\;
  \includegraphics[width=0.47\textwidth]{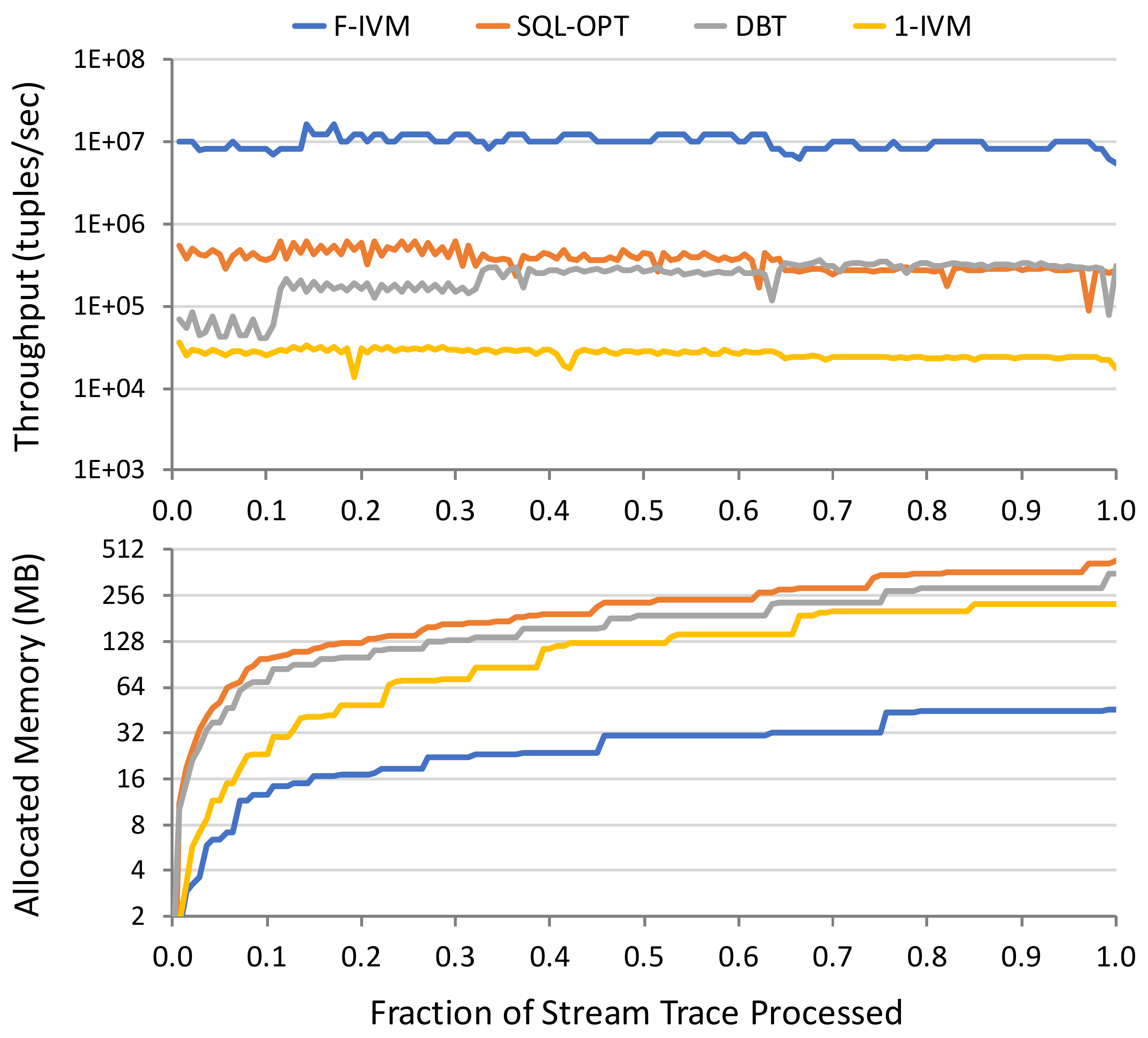}
  \caption{Incremental maintenance of the cofactor matrix over the {\em Retailer} dataset (left) and {\em Housing} dataset (right) under updates of size $1,000$ to all relations with a one-hour timeout. The ONE plots consider updates to the largest relation only. }
  \label{fig:cofactor_IVM_trace_ALL}
\end{figure*}

\subsection{Cofactor Matrix Computation} 

We benchmark the performance of maintaining a cofactor matrix for learning regression models over a natural join. We compute the cofactor matrix over all variables of the join query (i.e., over all attributes of the input database), which suffices to learn linear regression models over {\em any label and set of features} that is a subset of the set of variables~\cite{OS:PVLDB:16}. This is achieved by specializing the convergence step to the relevant restriction of the cofactor matrix. In end-to-end learning of regression models over factorized joins in the {\em Retailer} and {\em Housing} datasets, the convergence step takes orders of magnitude less time compared to the data-dependent cofactor matrix computation~\cite{SOC:SIGMOD:2016}.

In addition to the three incremental strategies from before, we now also benchmark 
\DBTRING, DBToaster's recursive IVM strategy with payloads from the degree-$m$ ring (cf.~Section~\ref{sec:application-lr}) instead of scalars, 
and \SQLOPT, an optimized SQL encoding of cofactor matrix computation. 
The latter arranges regression aggregates -- recall there are quadratically many such aggregates in the number of query variables -- into a {\em single} aggregate column indexed by the degree of each query variable. 
This encoding takes as input a variable order and constructs one SQL query that intertwines join and aggregate computation by pushing (partial) regression aggregates (counts, sums, and cofactors) past joins~\cite{Olteanu:FactorizedDB:2016:SIGREC}.

We consider updates to all relations in the {\em Retailer} and {\em Housing} datasets. 
In the {\em Retailer} schema, \DF and \SQLOPT rely on a given variable order.
These two strategies store $9$ views each: five views over the input relations, three intermediate views, and the top-level view; \DBTRING stores four additional views, $13$ in total. These views are identical to those used for maintaining a sum aggregate but have different payloads.
DBToaster's recursive higher-order IVM and first-order IVM use scalar payloads and fail to effectively share the computation of regression aggregates, materializing linearly many views in the size of the cofactor matrix: \DBT and \IVM use $3,814$ and respectively $995$ views to maintain $990$ aggregates.
In the {\em Housing} schema, where all relations join on one variable, \DF and \SQLOPT materialize one view per relation and the root view, $7$ in total, while \DBT and \IVM use $702$ and respectively $412$ views to maintain $406$ aggregates.
\DF and \DBTRING use identical strategies for the {\em Housing} dataset.

Figure~\ref{fig:cofactor_IVM_trace_ALL} shows the throughput of these techniques as they process an increasing fraction of the stream of tuple inserts. 
The {\em Retailer} stream consists of inserts into the largest relation mostly, and since the variables of this relation form a root-to-leaf path in the variable order, processing a single-tuple update takes $\bigO{1}$ time for \DF and \SQLOPT. 
The former outperforms the latter due to efficient encoding of triples of aggregates $(\LRringC,\LRringS,\LRringQ)$ as payloads containing vectors and matrices. 
\DBTRING's additional views cause non-constant update times to the largest relation, which means $8.7$x lower average throughput than \DF.
The two approaches with scalar payloads, \DBT and \IVM, need to maintain too many views  and fail to process the entire stream within a one-hour limit.

The query for {\em Housing} is a star join with all relations joining on the common variable, which is the root in our variable order. Thus, \DF and \SQLOPT can process a single tuple in $\bigO{1}$ time. \DBTRING and \DF use the same strategy in this case.
\DBT exploits the conditional independence in the derived deltas to materialize each input relation separately such that all non-join variables are aggregated away. 
Although each materialized view has $\bigO{1}$ maintenance cost per update tuple, the large number of such views in \DBT is the main reason for its poor performance. 
In contrast, \IVM stores entire tuples of the input relations including non-join variables.
On each update, \IVM recomputes an aggregate on top of the join of these input relations and the update.
Since an update tuple binds the value of the common join variable, the hypergraph of the delta query consists of disconnected components. 
DBToaster optimizes such a delta query by placing an aggregate around each component, 
that is, the delta first aggregates over each relation and then joins together the partial aggregates.
Even with this optimization, \IVM takes linear time, which explains its poorer performance.


{\bf Memory Consumption.}
Figure~\ref{fig:cofactor_IVM_trace_ALL} shows that \DF achie\-ves the lowest memory utilization on both datasets while providing orders of magnitude better performance than its competitors! 
The reason behind the memory efficiency of \DF is twofold. 
First, it uses complex aggregates and factorization structures to express the cofactor matrix computation over a smaller set of views compared to \DBTRING and, even more, to \DBT and \IVM. 
Second, it encodes regression aggregates implicitly using vectors and matrices rather than explicitly using variable degrees, like in \SQLOPT. 
The occasional throughput hiccups in the plot are due to expansion of the underlying data structures used for storing views.

{\bf The Effect of Update Workload.}
Our next experiment studies the effect of different update workloads on performance. 
We consider the {\em Retailer} dataset and two possible update scenarios: (1) all relations can change, in which case every view in the view tree needs to be materialized; (2) only the largest relation changes, while all others are static (denoted as ONE in Figure~\ref{fig:cofactor_IVM_trace_ALL}). 
In the latter scenario, we can precompute the views that are unaffected by changes and avoid materialization of those views that do not directly join with the updated relation. Thus, restricting updates to only one relation leads to materializing fewer views, which in turn reduces the maintenance overhead. 
Figure~\ref{fig:cofactor_IVM_trace_ALL} shows the throughput of processing updates for  the incremental maintenance of the cofactor matrix in these two scenarios. If we restrict updates only to a relation, we can avoid materializing all the views on the leaf-to-root path covered by that relation.  
This corresponds to a streaming scenario where we compute a continuous query and do not store the stream. Restricting updates to only one relation improves the average throughput, $3.2$x in \DF and $1.3$x in \SQLOPT, and also decreases memory consumption (note the log $y$-axis).
The latter also reflects in smoother throughput curves for the ONE variants. 
In \DBT, restricting updates brings $\bigO{1}$ time updates per view, yet the number of views is still large.
\DF and \DBTRING use identical materialization strategies here.


\begin{figure*}[t]
\begin{minipage}[t]{.475\textwidth}
  \centering   
  \includegraphics[width=\columnwidth]{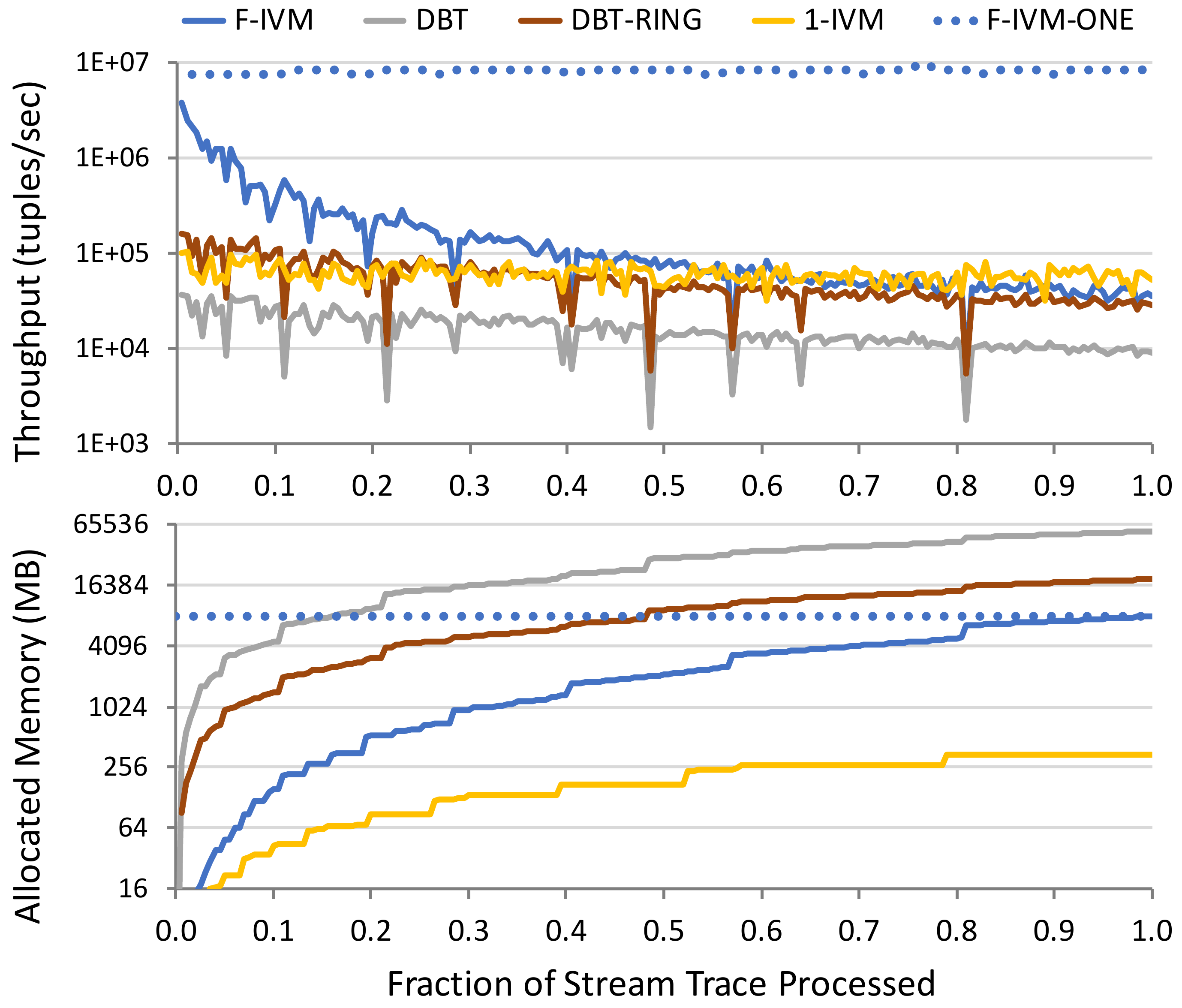}
  \caption{Incremental maintenance of the cofactor matrix on top of the triangle query on {\em Twitter} for updates of size $1,000$ to all input relations.}
  \label{fig:cofactor_Triangle_IVM_trace_ALL}
\end{minipage}
\quad
\begin{minipage}[t]{.475\textwidth}
  \centering   
  \includegraphics[width=\columnwidth]{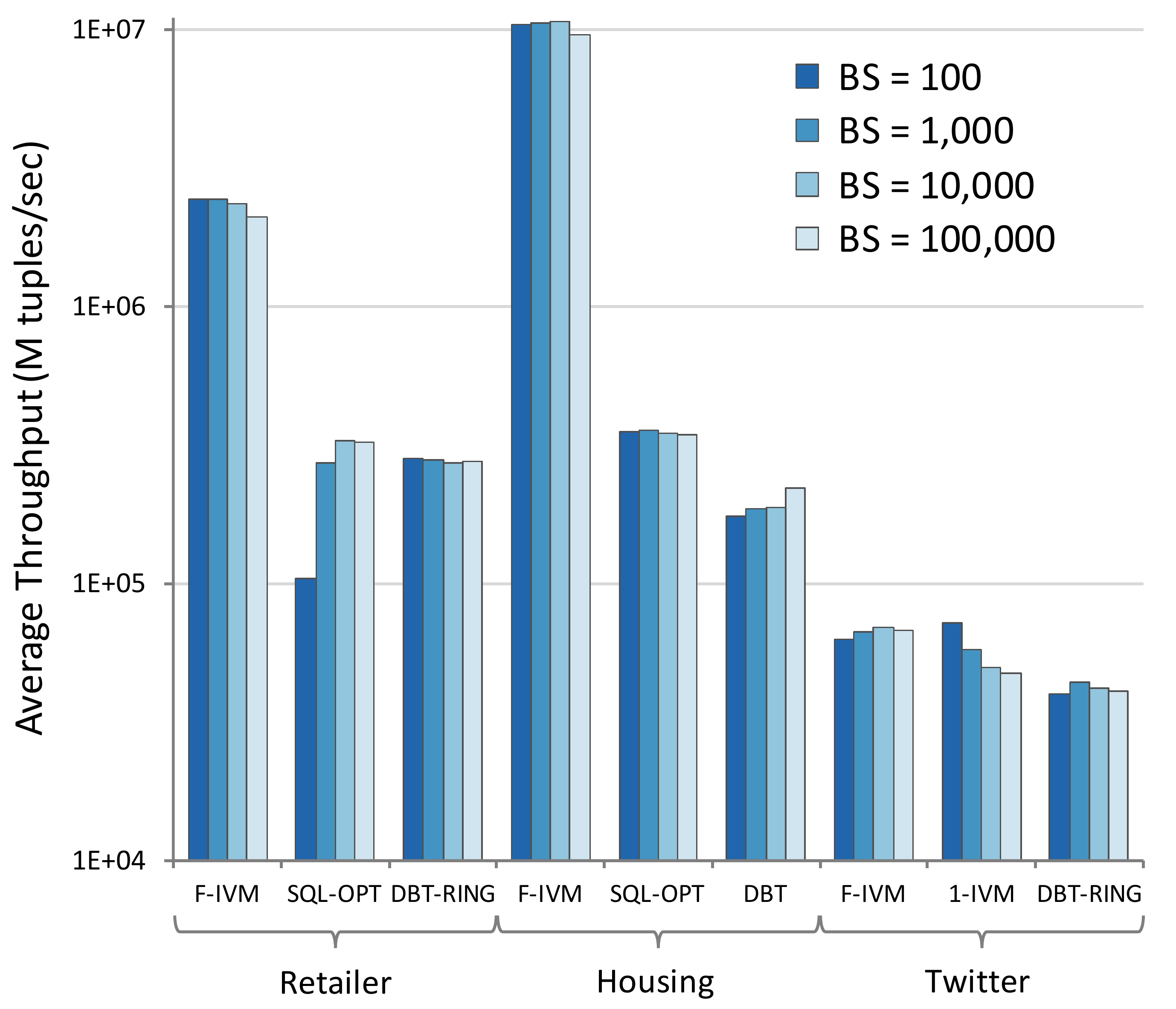}
  \caption{Incremental maintenance of the cofactor matrix under batch updates of different sizes to all input relations. }
  \label{fig:cofactor_IVM_batch_sizes_ALL}
\end{minipage}%
\end{figure*}


{\bf Cofactor Matrix Computation over the Triangle Query.}
We analyze the cofactor matrix computation over the triangle query on the {\em Twitter} dataset and updates of size $1,000$ to all the relations.
\DF uses the view tree from Figure~\ref{fig:triangle_hypergraph_viewtree}~(right) without the indicator projection and materializes the join of $S$ and $T$ of size $\bigO{N^2}$. 
Its time complexity for a single-tuple update to $R$ is $\bigO{1}$, but updating the join of $S$ and $T$ takes $\bigO{N}$. 
\DBTRING uses payloads from the degree-$3$ ring and materializes all three such pairwise joins, each requiring linear time maintenance. 
\DBT uses scalar payloads and materializes $21$ views (to maintain $6$ aggregates), out of which $12$ views are over two relations. Its time complexity for processing single-tuple updates to either of the three relations is also $\bigO{N}$.
The \IVM strategy maintains just the input relations and recomputes the delta upon each update in linear time.

The throughput rate of the strategies that materialize views of quadratic size declines sharply as the input stream progresses. \DBT exhibits the highest processing and memory overheads caused by storing $12$ auxiliary views of quadratic size. \DBTRING underperforms \DF due to maintaining two extra views of quadratic size, which contribute to a $2.3$x higher peak memory utilization. \IVM exhibits a $42$\% decline in performance after processing the entire trace due to its linear time maintenance. The extent of this decrease is much lower compared to the other approaches with the quadratic space complexity. For updates to $R$ only, \DFONE (which is identical to \DBTRING's ONE variant) requires one lookup in the materialized join of $S$ and $T$ per update. This strategy has two orders of magnitude higher throughput than \IVM at the cost of using $23$x more memory.

Clique queries like triangles provide no factorization opportunities. Materializing auxiliary views to speed up incremental view maintenance increases memory and processing overheads. However, \DF can exploit indicator projections to bound the size of such materialized views, as described in Section~\ref{sec:cyclic_queries}.



{\bf The Effect of Batch Size on IVM.}
This experiment evaluates the performance of maintaining a cofactor matrix for batch updates of different sizes. Figure~\ref{fig:cofactor_IVM_batch_sizes_ALL} shows the throughput of batched incremental processing for batch sizes varying from $100$ to $100,000$ on the {\em Retailer}, {\em Housing}, and {\em Twitter} datasets for updates to all relations.
We show only the best three approaches for each dataset.

We observe that using very large or small batch sizes can have negative performance effects: Iterating over large batches invalidates previously cached data resulting in future cache misses, whereas using small batches cannot offset the overhead associated with processing each batch. 
Using batches with $1,000-10,000$ tuples delivers best performance in most cases, except when needed to incrementally maintain a large number of views. This conclusion about cofactor matrix computation is in line with similar findings on batched delta processing in decision support workloads~\cite{Nikolic:Batching:2016:SIGMOD}.

Batched incremental processing is also beneficial for one-off computation of the entire cofactor matrix. Using medium-sized updates can bring better performance, cf.\@ Figure~\ref{fig:cofactor_IVM_batch_sizes_ALL}, but can also lower memory requirements and improve cache locality during query processing. 
For instance, incrementally processing the {\em Retailer} dataset in chunks of $1,000$ tuples can bring up to $2.45$x better performance compared to processing the entire dataset at once.

\begin{figure*}[t]
  \centering   
  \includegraphics[width=0.48\textwidth]{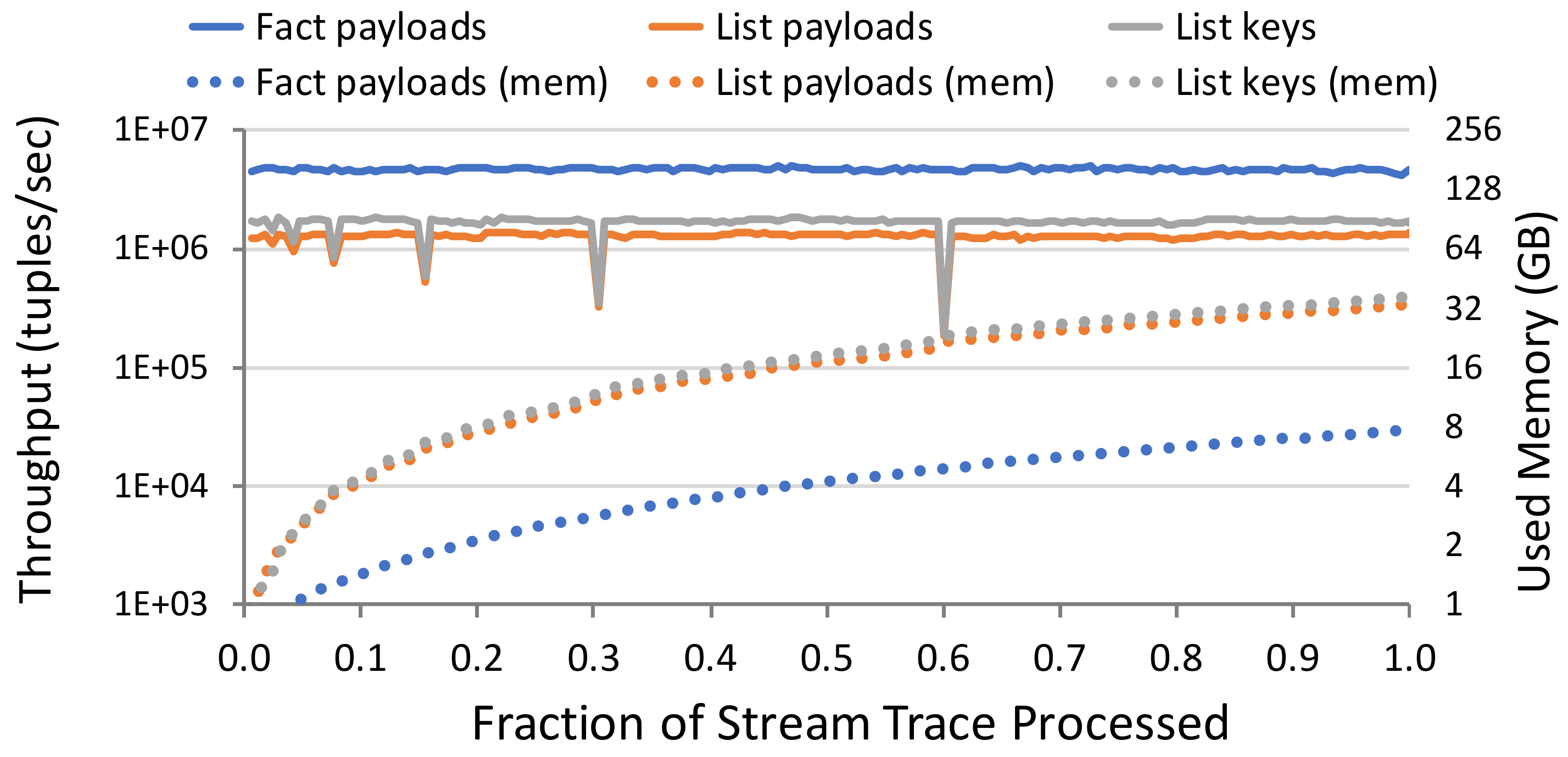}
  \quad
  \includegraphics[width=0.48\textwidth]{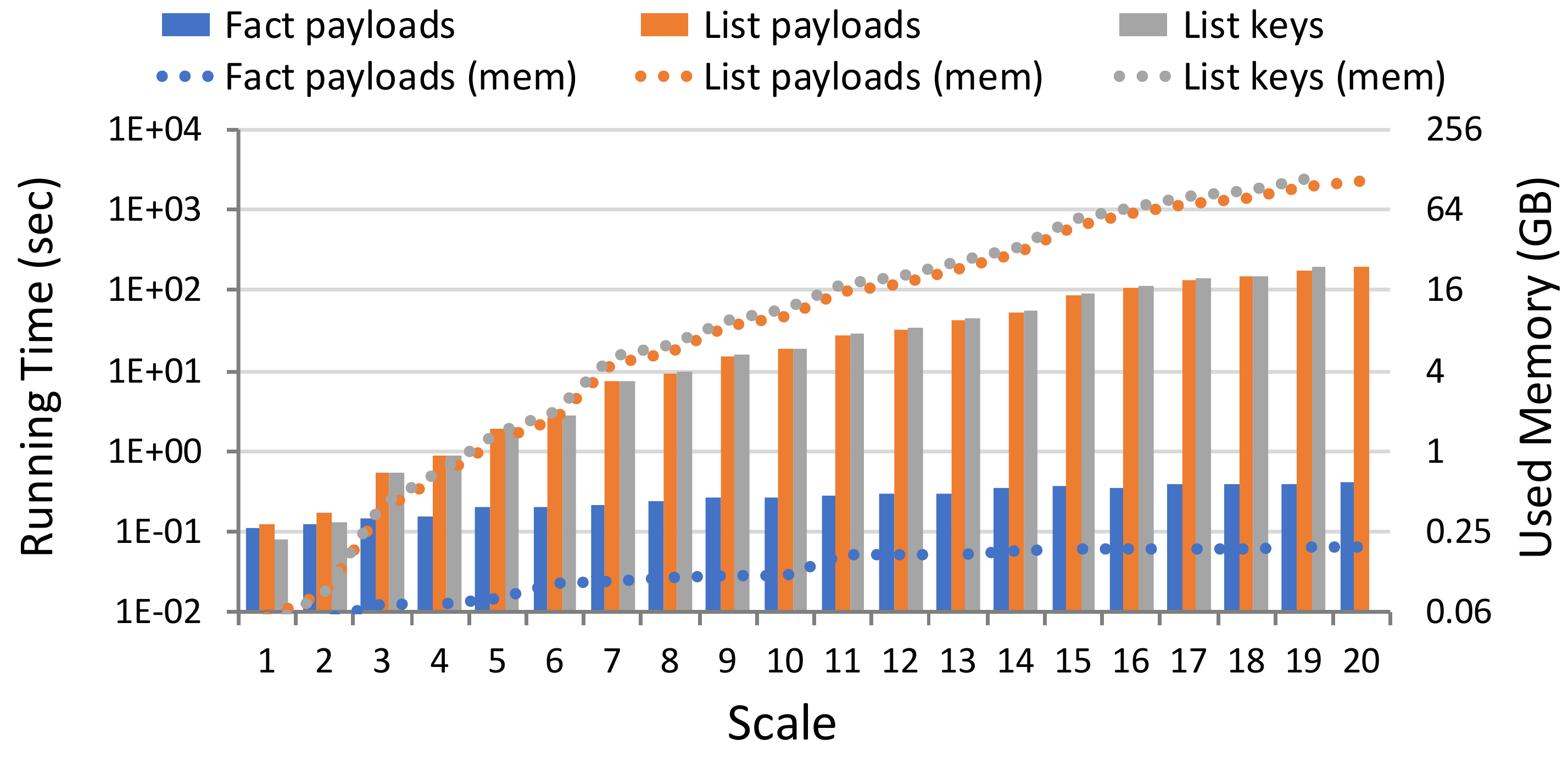}
  \caption{Incremental maintenance using relational and factorized payloads for the natural joins of the {\em Retailer}  (left) and of the {\em Housing} (right) datasets under updates of size $1,000$ to the largest relation ({\em Retailer}) and all input relations ({\em Housing}).}
  \label{fig:FullJoin_Factorized_Relational}
\end{figure*}

\subsection{Factorized Computation of Conjunctive Queries}

We analyze \DF on queries whose results are stored as keys with integer multiplicities using listing representation ({\tt List$\;$keys}) and as relational payloads using factorized and listing representations ({\tt Fact$\;$payloads} and {\tt List$\;$payloads}).
Figure~\ref{fig:FullJoin_Factorized_Relational} (left) considers the natural join of {\em Retailer} under updates to the largest relation. The factorized payloads reduce the memory consumption by $4.4$x, from $34$GB to $7.8$GB, improve the average throughput by $2.8$x and $3.7$x (and the overall run time by $3.2$x and $4.2$x) compared to using the two listing encodings.
Figure~\ref{fig:FullJoin_Factorized_Relational} (right) considers the natural join of {\em Housing} under updates to all input relations.  The number of tuples in the dataset varies from $150,000$ (scale 1) to $1,400,000$ (scale 20), while the size of the listing (factorized) representation of natural join grows cubically (linearly) with the scale factor. The two listing encodings blow up the memory consumption and computation time for large scales. Storing tuples in the listing representation using payloads instead of keys avoids the need for hashing wide keys, which makes the joins slightly cheaper. For {\em Housing} and factorized representation, the root view stores $25,000$ values of the join variable regardless of the scale. The root's children map these values to relational payloads for each relation. For the largest scale, {\tt Fact$\;$payloads} is $481$x faster and takes $548$x less memory than {\tt List$\;$payloads} ($410$ms vs. $197$s, $195$MB vs. $104$GB), and {\tt List$\;$keys} exceeds the available memory.

\nop{In this setting, DBToaster's recursive IVM is equivalent to the {\tt List$\;$keys} strategy.}

\nop{The compression effect is significant but limited due to the key constraints in {\em Retailer}. 
The hash maps used for storing the views maintain memory pools of increasingly larger sizes, which reflects in occasional throughput hiccups that correspond to the expansion of the underlying data structures. }


\section{Related Work}
\label{sec:related_work}

To the best of our knowledge, ours is the first approach to propose factorized IVM for a range of distinct applications. It extends non-trivially two lines of prior work: higher-order delta-based IVM and factorized computation of in-database analytics. 

Our view language is modeled after functional aggregate queries over semirings~\cite{FAQ:PODS:2016} and generalized multiset relations over rings~\cite{DBT:VLDBJ:2014}; the latter allowed us to adapt DBToaster to factorized IVM.

{\bf IVM.} IVM is a well-studied area spanning more than three deca\-des~\cite{Chirkova:Views:2012:FTD}. Prior work extensively studied IVM for various query languages and showed that the time complexity of IVM is lower than of recomputation. We go beyond prior work on higher-order IVM for queries with joins and aggregates, as realized in DBToaster~\cite{DBT:VLDBJ:2014}, and propose a unified approach for factorized computation of aggregates over joins~\cite{BKOZ:PVLDB:2013}, factorized incremental computation of linear algebra~\cite{NEK:SIGMOD:2014}, and learning regression models over factorized joins~\cite{SOC:SIGMOD:2016}. DBToaster uses one materialization hierarchy per relation in the query, whereas we use one view tree for all relations. DBToaster can thus have much larger space requirements and update times. DBToaster does not target the maintenance of many complex aggregates that share computation (e.g., cofactor matrices), which we observe experimentally. IVM over array data~\cite{Zhao:2017:ArrayIVM} targets scientific workloads but without exploiting data factorization.

Our approach with the relational payload ring strictly subsumes previous work on factorized IVM for acyclic joins~\cite{DynYannakakis:SIGMOD:2017}
as it can support cyclic joins (see Section~\ref{sec:cyclic_queries}).
The so-called $q$-hierarchical join queries (such as the Housing query in our experiments) are exactly those self-join-free conjunctive queries that admit constant time update~\cite{Nicole:PODS:2017}. 
Recent work on in-database maintenance of linear regression models shows how to compute such models using previously computed models over distinct sets of features~\cite{Gupta:CORR:2015}. Its contribution is complementary to ours and shares a similar goal with prior work on reusing gradient computation to efficiently explore the space of possible regression models~\cite{OS:PVLDB:16}. Exploiting key attributes to enable succinct delta representations and accellerate maintenance complements our approach~\cite{Katsis:idIVM:2015}.
Our framework generalizes the main idea of the LINVIEW approach~\cite{NEK:SIGMOD:2014} to maintenance of matrix computation over arbitrary joins. 

Most commercial databases, e.g., Oracle~\cite{Oracle:RestrictionsIVM} and SQLServer~\cite{SQLServer:RestrictionsIVM}, support IVM for restricted classes of queries. LogicBlox supports higher-order IVM for Datalog (meta)programs~\cite{LB:SIGMOD:2015,GOW:PVLDB:2015}. Trill is a streaming engine that supports incremental processing of relational-style queries but no complex aggregates like cofactor matrices~\cite{chandramouli2014trill}. 

{\bf Static In-DB analytics.} The emerging area of in-data\-base analytics has been recently overviewed in two tutorials~\cite{Polyzotis:SIGMOD:Tutorial:17,Kumar:SIGMOD:Tutorial:17}.  Several systems support complex analytics over normalized data via a tight integration of databases and machine learning~\cite{MLlib:JMLR:2016,MADlib:2012,Rusu:2015,Polyzotis:SIGMOD:Tutorial:17,Kumar:SIGMOD:Tutorial:17}. Others integrate with R to enable in-situ data processing using domain-specialized routines~\cite{ZCDDMMFSS12,Brown:SciDB:2010:SIGMOD}. The closest in spirit to our approach is work on learning models over factorized joins~\cite{Rendle13,SOC:SIGMOD:2016,OS:PVLDB:16, ANNOS:PODS:2018}, pushing ML tasks past joins~\cite{Kumar:InDBMS:2012} and on in-database linear algebra~\cite{Boehm:VLDB:2016,Arun:VLDB:2017}, yet they do not consider incremental maintenance.

{\bf Learning.} There is a wealth of work in the ML community on incremental or online learning over {\em arbitrary} relations~\cite{OnlineML:2011}. Our approach learns over {\em joins} and crucially exploits the join dependencies in the underlying training dataset to improve the performance. 

\nop{
Our framework can be used for learning regression models over joins, which follows a recent line of research on marrying databases and machine learning~\cite{MADlib:2012,CMPW13,BTRSTBV14, KuNaPa15,MLlib:JMLR:2016,Neumann15,HBTRTR15,CGLPVJ14,SSMBRE15,RABCJKR15,Rusu:2015} and in particular builds on {\em static} factorized in-database learning~\cite{Kumar:InDBMS:2012,Rendle13,SOC:SIGMOD:2016,OS:PVLDB:16}. Our factorization approach is that from prior work~\cite{SOC:SIGMOD:2016,OS:PVLDB:16}. Limited forms of factorized learning have been also used by Rendle~\cite{} and Kumar et al.~\cite{KuNaPa15}. The former considers zero-suppresed design matrices for high-degree regression models called factorization machines. The latter proposes a framework for learning generalized linear models over key-foreign key joins in a distributed environment.
}
\nop{
Most efforts in the database community are on designing systems to support large-scale machine learning libraries on distributed architectures~\cite{Kumar:InDBMS:2012}, e.g., MLLib~\cite{MLlib:JMLR:2016} and DeepDist~\cite{Neumann15} on Spark \cite{ZCDDMMFSS12}, GLADE~\cite{Rusu:2015}, MADlib \cite{MADlib:2012} on PostgreSQL, SystemML \cite{HBTRTR15,BTRSTBV14}, system benchmarking~\cite{CGLPVJ14} and sample generator for cross-validate
learning~\cite{SSMBRE15}. 
}


\section{Conclusion}

This paper introduces a unified IVM approach for analytics over dynamic normalized data and shows its applicability to three seemingly disparate analytical tasks: matrix chain multiplication, query evaluation with listing/factorized result representation, and gradient computation used for learning linear regression models. These tasks use the same computation paradigm that factorizes the representation and the computation of the keys, the payloads, and the updates. Their differences are factored out in the definition of the sum and product operations in a suitable ring. This approach has been implemented as an extension of DBToaster, a state-of-the-art system for incremental maintenance, and shown to outperform competitors by up to two orders of magnitude in both time and space. 

Going forward, we would like to apply this approach to further tasks such as inference in probabilistic graphical models and more complex machine learning tasks.

\paragraph*{Acknowledgments.}
This project has received funding from the European Union's Horizon 2020 research and innovation programme under grant agreement No 682588. The authors also acknowledge awards from Microsoft Azure via The Alan Turing Institute and from Fondation Wiener Anspach.

\vspace{1em}
\begin{quote}{\em \hspace*{-1em} If You Liked It, Then You Should Put A Ring On It.\\
\hspace*{14.5em} -- Beyonc\'e.}
\end{quote}

\bibliographystyle{abbrv}
\bibliography{bibliography}

\end{document}